\documentclass[a4paper,12pt]{article}
\usepackage{a4}
\usepackage{amsfonts,amssymb,amsmath}
\usepackage{graphicx}
\begin{document}

\begin{center}
\Large \bf Interaction of a rotational motion and an axial flow in small
geometries for a Couette-Taylor problem
\end{center}
\begin{center} {
L. A. Bordag${}^*$\footnote{e-mail: ljudmila@bordag.com, bordag@math.tu-cottbus.de},
O. G. Chkhetiani${}^\dag$\footnote{e-mail: ochkheti@mx.iki.rssi.ru},
M. Fr{\"o}hner${}^*$\footnote{e-mail: froehner@math.tu-cottbus.de} and
V. Myrnyy${}^*$}\footnote{e-mail: myrnyy@math.tu-cottbus.de}\\[5pt]
${}^*$ {\it Fakult{\"a}t Mathematik, Naturwissenschaften und Informatik\\
  Brandenburgische Technische Universit{\"a}t Cottbus\\
  Universit{\"a}tsplatz 3/4, 03044 Cottbus, Germany}\\
${}^\dag${\it Space Research Institute,
Russian Academy of Sciences, \\ Profsoyuznaya 84/32,117997  Moscow }\\[5pt]
\end{center}

\begin{abstract}
We analyze the stability of a cylindrical Couette flow under the
imposition of a weak axial flow in case of a very short cylinder with
a narrow annulus gap.

We consider an incompressible viscous fluid which is contained in the
narrow gap between two concentric short cylinders, where the inner
cylinder rotates with constant angular velocity. The caps of the
cylinders have narrow tubes conically tapering to super narrow slits
which allow for an axial flow along the surface of the inner cylinder.

The approximated solution for the Couette flow for short cylinders was
found and used for the stability analysis instead of the exact but
bulky solution.  The sensitivity of the Couette flow to general small
perturbations and to the weak axial flow was studied. We demonstrate that
perturbations coming from the axial flow cause the propagation of
dispersive waves in the Taylor-Couette flow.

The coexistence of a rotation and of an axial flow requires to study
in addition to the energy and the angular momentum also the helicity
of the flow.  The approximated form for
the helicity formula in case of short cylinders was derived.

We found that the axial flow stabilizes the Taylor - Couette flow. The
supercritical flow includes a rich variety of vortical structures
including a symmetric pair of Taylor vortices, an anomalous single
vortex and quasi periodic oscillating vortices.  Pattern formation was
studied at large for rated ranges of azimuthal and axial Reynolds
numbers.  A region where three branches of different states occur was
localized.  Numerical simulations in 3D and in axisymmetrical case of
the model flow are presented, which illustrate the instabilities
analyzed.

\end{abstract}


\section{Introduction}\label{intr}
Since the famous experiments of Taylor \cite{tay} in long cylinders
many experimental and theoretical investigations of the interesting
phenomena of arising and evolution of Taylor vortices were done. The
main part of these studies was devoted to cases of long cylinders. It
was done under the assumption that a sufficiently long cylinder and
periodic boundary conditions will emulate the infinitely long
cylindrical region well.  In addition such regions are convenient for
theoretical investigations. The experimental investigations of short
cylinders started much later and gave surprising results (see, for
example, Benjamin and Mullin \cite{mb}), which indicated that the zone
and the magnitude of the influence of the caps of the cylinders is
much bigger and takes effect on the type of motion in the whole
region.  It was assumed that close to the caps of the cylinders the
boundary layer is directed from the outer to the inner cylinder as a
result of the existence of an Ekmans boundary layer.  However, the
experiments gave a more complicated picture of the motions. There
exists also an atypical layer in opposite direction, i.e. from the
inner to the outer cylinder connected withe the existence of the 'anomalous mode'. The
first mathematical model which took the influence of the cylinder caps
into account was represented in the work of Schaeffer \cite{sch}. This
model is not very consistent and needs improvements. The role of the
anomalous mode was studied later numerically and experimentally in the
works of Cliffe et al \cite{cm}, \cite{ckm}. The transient features of
the circular Couette flow were investigated in laboratory and
numerical experiments (using the commercial program {\it 'Nekton'}
(Fluent GmbH) by Neitzel et al \cite{nkl}. The results of the
numerical experiments were close to the laboratory experiments. In
case of short cylinders with an aspect ratio $\Gamma \sim 1$ (relation
of annulus span to the annulus gap width, defined in
section~\ref{main}) usually only one pair of Taylor vortices arises.
In general, this case is considered to be not really interesting due
to the poorness of the patterns to be expected.  But in the
experiments of Aitta et al \cite{aac} a series of interesting flow
patterns for short cylinders was found. A non-equilibrium tricritical
points occurs when a forward bifurcation becomes a backward
bifurcation. In the Taylor - Couette problem the tricritical
point can be observed for $\Gamma_T = 1.255$  in the
classical case of a pure rotation of the inner cylinder without axial
flow as demonstrated in \cite{aac} and later in \cite{mu02}. The system of two vortices is symmetrical if the rotation
velocity does not exceed a critical velocity $v_1.$ If the velocity of
the flow is larger, $v>v_1$, then one vortex grows at the expense of
the other one. In the case of the forward bifurcation the symmetry is
broken continuously and in the case of the backward bifurcation -
abruptly. In the phase transition language this is a bifurcation of
the second order or first order correspondingly \cite{bhat}.  The
continuous transition was observed for $\Gamma < \Gamma_T $ and the
abrupt one for $\Gamma_T <\Gamma < \Gamma_C, $ where $ \Gamma_C =1.292
.$

An exhaustive study of the influence of the finite length effects is
done for the classical arrangement of two concentric cylinders with a
very small aspect ratio $\Gamma \in [0.5, 1.6]$ and a relatively large
annulus with a radius ratio (relation between the radii of inner and
outer cylinders, see section~\ref{main}) of $\eta = 0.667 $ for
moderate Reynolds numbers $Re \in [100, 1500]$ in the work of
Furukava, Watanabe, Toya \& Nakamura \cite{furuwa} in the case when
the inner cylinder is rotating. The numerical investigations were
verified by experiments. The three main flow patterns were found which correspond to a
normal two-cell mode, an anomalous one-cell mode and a twin - cell
mode. A steady case as well as an unsteady mode of the fully developed
flow different from wavy Taylor - Couette flow were obtained.

There exists a series of experimental and theoretical works devoted to
the linear stability of the Taylor-Couette problem with implosed axial
effects. The first such analysis was done for the case of axisymmetric
disturbances in a narrow gap \cite{Chandra}, \cite{DiPrima}.  In the
work of Marques and Lopez \cite{mar} the linear stability of the flow
in the annulus between two infinitely long cylinders, driven by a
constant rotation and harmonic oscillation in the axial direction of
the inner cylinder was analyzed using Floquet theory.  The axial
effects in the Taylor-Couette problem were studied in the work
\cite{mesmar} on two different flows. In the first one the axial
effect is introduced by an inertial axial sliding mechanism between
the cylinders and in the other one via an imposed axial pressure
gradient. In all cases the annular gap is much larger as in our case
and the radius ratio is $\eta=0.5,~\dots,~0.8.$ The stabilizing effect
of a periodical axial flow was studied in the works \cite{Hu},
\cite{cha} and \cite{weis}. In the work of Lueptow \cite{luep} the
stability problem of a Taylor-Couette flow with axial and radial flow
was studied for a relatively large radius ratio $\eta=0.83$ and
pressure-drive axial flow in the whole annulus.  In all these studies
it was found that the axial flow stabilized the Taylor-Couette flow
and enables a lot of super critical states with a rich variety of
patterns.

Nevertheless the theory of the Taylor - Couette flow is by far from
being complete, so for instance the stability problem for the system
of Taylor vortices and a classical Couette flow by influencing control
of the cylinder caps.  It is also a little bit astonishing that the
exact analytical solution for the Couette flow profile with boundary
conditions on cylinder caps was obtained only recently in the work of
Wendl \cite{wen}. His results expose a strong influence of the caps on
the flow profile. The difference between the two profiles with and
without finite-length effects is approximately a logarithmic function
of the aspect ratio $\Gamma$ for a wide range of radius ratios
$\eta$. The solution is given in form of a slowly convergent series
which contains trigonometric and Bessel functions. To give a
qualitative analysis of instabilities one takes an exact solution as a
base flow to linearize the Navier - Stokes equations. This type of
solutions is not convenient to pursue the topic further towards
analytical studies of the instabilities.

In the present work we use the theory of small perturbations
to study the instability of the laminar Couette flow in a short
cylinder.  The geometry of an annular gap and the boundary conditions
are given in the section~\ref{main}.  Our idea was to get an
analytical formula which is simpler than that in \cite{wen} but which
gives us an approximation of the Couette flow with finite-length
effects which is as good as that found in \cite{wen}. In the
section~\ref{sero} we describe this solution and discuss the region of
applicability of the done approximation.

The solution obtained for a short cylinder is well suitable as a base
flow for a comprehensive study of instability effects. This study is
represented in the section~\ref{loc}. The onset of the instability is
predicted using the method of local analysis, in which
the equations governing the flow perturbation are linearized. For the
very reason that the appearance of an instability depends not only on
the base flow but also on the amplitude and the type of the flow
perturbation two different types of perturbations are studied in this
section.  We consider the sensitivity of the Couette flow under
small perturbations corresponding to an axial flow which is
the most important for industrial applications. We take the case of a
very narrow annulus $\eta \sim 1$ which is typical for bearing or
sealing systems and in contrary to general expectation we get a rich
family of flow states.  We proved that the axial flow stabilizes the
Taylor - Couette flow also in case of very short cylinders and narrow
annular gap with axial flow directed as a thin layer along the surface
of the inner rotating cylinder.

The elaborate series of the numerical experiment was done with help of
the FLUENT 5.5/6.0. The detailed description of the grids used and
others numerical facilities is given in the section~\ref{sec7}.

The change from one laminar flow to another or a transition from a
laminar flow to turbulence is a function of the base flow considered
as well as of the type of small perturbations.  We found a
rich family of typical patterns on large intervals of Reynolds
numbers.  The results of the numerical experiments are discussed in
the section~\ref{resul}.

\section{The main problem}\label{main}

The study of the Couette - Taylor problem for the parameters defined
 below was stimulated by the industrial project \cite{bor}.  In many
 technical systems the problem of transmitting energy and momentum is
 solved using shafts and corresponding bushes with or without radial
 sealing. In a typical situation one is concerned with a small
 Taylor-Couette annulus, narrow slits in between cylinders and with
 axial and tangential oscillations or with a weak axial flow along the
 surface of the inner cylinder. In general, these systems are very
 complicated and it is necessary to study more simple problems in
 order to isolate a particular mechanism and to develop and test
 hypotheses governing their behavior.  We separate here only one of
 the most typical cases because of its technological importance and,
 from a fundamental point of view, for its rich dynamics and try to
 describe all possible flow states in this case.

\begin{figure}[ht]
\includegraphics[width=10cm]{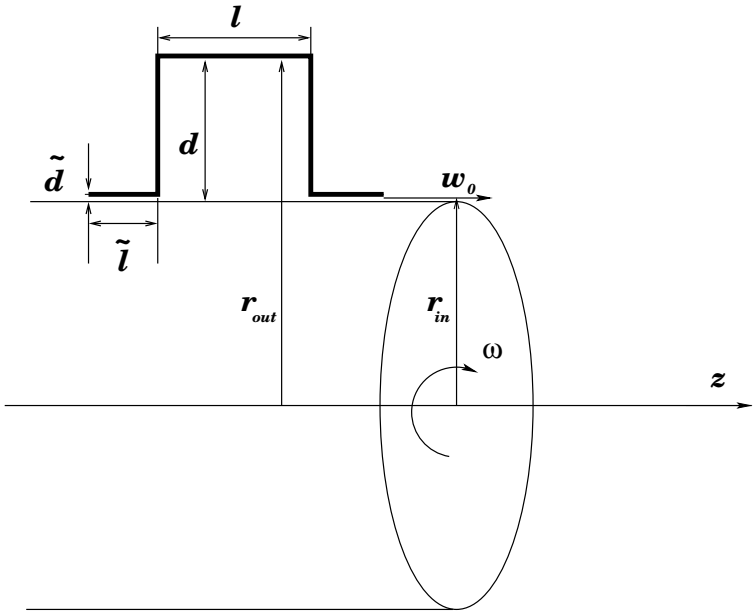}
\caption{\label{k2} \small The geometry of the annulus. The radii of the inner and outer cylinders
  are equal to $r_{in}, r_{out}$ correspondingly, the annulus span is
  denoted by $l,$ the annular gap width is $d=r_{out}-r_{in}.$ The
  corresponding geometrical parameters of the slits are denoted by
  $\tilde d$ and $\tilde l.$ Other notations are explained in the
  text.}
\end{figure}

We consider the motion of the fluid in an annulus between two cylinders as
represented in fig.~\ref{k2}.
The annulus dimensions are the gap width $d$, the annulus span $l$,
the radius of the inner cylinder $r_{in}$, the of an outer cylinder
$r_{out}$.  The corresponding dimensionless geometric parameters are
the aspect ratio $\Gamma = \frac{l}{d}$ and the radius ratio
$\eta=\frac{r_{in}}{r_{out}}$. We use these notations during the whole
paper.  For many sealing or bearing systems $\Gamma \sim 1$ and $\eta
\sim 1.$ We fixe in all our calculations $\Gamma =1.02$, $
\eta=0.95356~.$ 

We assume that the inner cylinder to turn at the constant rate
$\omega .$ The outer cylinder is at rest, the axial flow has a constant
velocity $w_0$ and we suppose $r _{in}\omega \gg w_0.$ The equations
of fluid motion and boundary conditions will be rendered dimensionless
using $d,$ as a unit for length. The radius of the inner cylinder will
be denoted by $R=\frac{r_{in}}{d}$ in dimensionless units which will
be used in all sections of this paper.

The entrance way and the outgoing way for the axial flow are very thin
cylindrical slits with the gap width $\tilde{d}$ and the annulus span
of slits denoted by $\tilde{l}.$ The slits are long and narrow in
comparison to the main annulus i.e. $l \sim \tilde{l},$ $d \gg \tilde
d.$ The ends of the slits narrow conical to the lower order of scale
that means the most narrow part of the slit has the width
$\widetilde{\widetilde d},$ which is very small, $d \gg \tilde d \gg
\widetilde{\widetilde d} $. A more detailed picture of the slits is
given in fig.~\ref{2dgrid}.  The studied flow states are very
sensitive to the kind of perturbations. So we try to construct a
situation which is most adapted to technical facilities and imitate
the sealing arrangements with conical tapering slits.

It is characteristically for quite all technical applications that the
sealing medium is oil with non vanishing viscosity.  We will consider
in our work an incompressible viscous fluid in the cylindrical
annulus.

\section{An approximation of the exact solution in case of the rotating inner
  cylinder}\label{sero}

The classical case of the Couette flow of an incompressible viscous
fluid in an annulus between two cylinders was investigated
analytically in \cite{wen}.  In comparison to the famous work of
Taylor the author studied cylinders of a finite length and got a much
more complicated formula which describes the influence of the cylinder
caps on the flow.  As follows from experiments of Kageyama (\cite{kag})
if the Reynolds number grows the influence of the Ekmans layer drops
along the cylinder caps. As a result the profile of the solutions
differs more and more from the pure Couette solution because of the
arising of secondary flows in the annulus.

We will consider the flow in the annular gap with the width $d,$ the
span $l,$ and radius of inner cylinder $r_{in}$ as on the
fig.~\ref{k2} without slits.  We suggest now that the caps of the
cylinders contact the inner cylinder and that there is no axial flow
in the annulus.  The equations of the fluid motion and the boundary
conditions will be rendered dimensionless using $v= r_{in} \omega$ and
$P=(\nu /d)^2$ as units for velocity and the pressure.  The scaling of
the velocities and of other quantities will change from section to
section.

Due to the axial symmetry two of the components of the velocity vector
${\bf u}(r,\theta,z)$ vanish in this case and the velocity vector
reads
\begin{equation}
{\bf u}(r,z)=(0,V_0(r,z),0),
\end{equation}
where $V_0(r,z)$ is the azimuthal component. The cylindrical
Navier-Stokes equations can be reduced in this model to one equation,
\begin{equation}
\frac{\partial ^2 V_0}{\partial r ^2}+ \frac{1}{R+r}\frac{\partial  V_0}{\partial r}
- \frac{V_0}{(R+r)^2} +\frac{1}{\Gamma^2}\frac{\partial ^2 V_0}{\partial z ^2}=0.\label{eqv}
\end{equation}
The boundary conditions can be rendered dimensionless as
\begin{equation}
V_0(r,0)=0,~~V(r, 1)=0,~~V_0(0,z)=1,~~V_0(1,z)=0,~z\in[0,1],~r\in[0,1]. \label{bcv}\\
\end{equation}
We make an additional scaling in $z-$direction by $\Gamma$ to reach
the more convenient situation that the both variables $r$ and $z$ lie
in the interval $[0,1].$ The considered case was solved by Wendl
\cite{wen} using the Fourier transformation.  The solution has the
following form
\begin{align}
&V_{0}(r,z)= \label{sov}\\
&\frac{1}{4\pi}  \sum_{m=1}^{\infty}{\frac{\left(
I_{1}(\beta_{m}(R+1))K_{1}(\beta
_{m}(R+r))-K_{1}(\beta_{m}(R+1))I_{1}(\beta_{m}(R+r))\right)
\sin(\beta _{m}z)}{(2m-1)\left(
I_{1}(\beta_{m}(R+1))K_{1}(\beta_{m}R)-K_{1}(\beta
_{m}(R+1))I_{1}(\beta_{m}R)\right)  }}\nonumber
\end{align}
with $ \beta_m=\frac{(2 m -1) \pi}{ \Gamma }$.  Here $I_1$ and $K_1$
are modified Bessel functions.  This solution differs from Taylor-Couette's
simple linear solution especially strong in cases of short cylinders but
remains an exact solution for long cylinders.  The series in
(\ref{sov}) converges very slowly and it is difficult to use it for
the intended analytical studies. This was the reason that we looked
for simpler analytical representations of this solution or for some
good approximated solution.

We study the case of short cylinders, i.e. $\Gamma \sim 1,$ with a
narrow gap, $\eta \sim 1$, and $R \gg 1.$ Under these assumptions and
taking into account $r\in [0,1]$ we can simplify the initial equation
(\ref{eqv}) by replacing $\frac {1}{R+r},\frac{1}{(R+r)^{2}}$ by the
constants $\frac{1}{R},\frac {1}{R^{2}}.$ For the simplified equation
which has constant coefficients we obtain the solution in closed
analytical form,
\begin{equation}
V_0(r,z)={\frac{2}{\pi}}\arctan \left(\left(
\sinh^{-1}\left(\frac{\pi} {\Gamma}r
\right)-\sinh^{-1}\left(\frac{\pi~}{\Gamma}\left( 2 -r \right)
\right)\right) \sin(\pi z)\right)\label{appsov}
\end{equation}
for $r\in [0,1]$ and $z \in [0,1]$.  This is a convenient form of the
approximated solution for the fundamental mode of the Couette flow
profile which we will use in the following.

\begin{figure}[ht]
\begin{minipage}[t]{7cm}
\includegraphics[width=7cm]{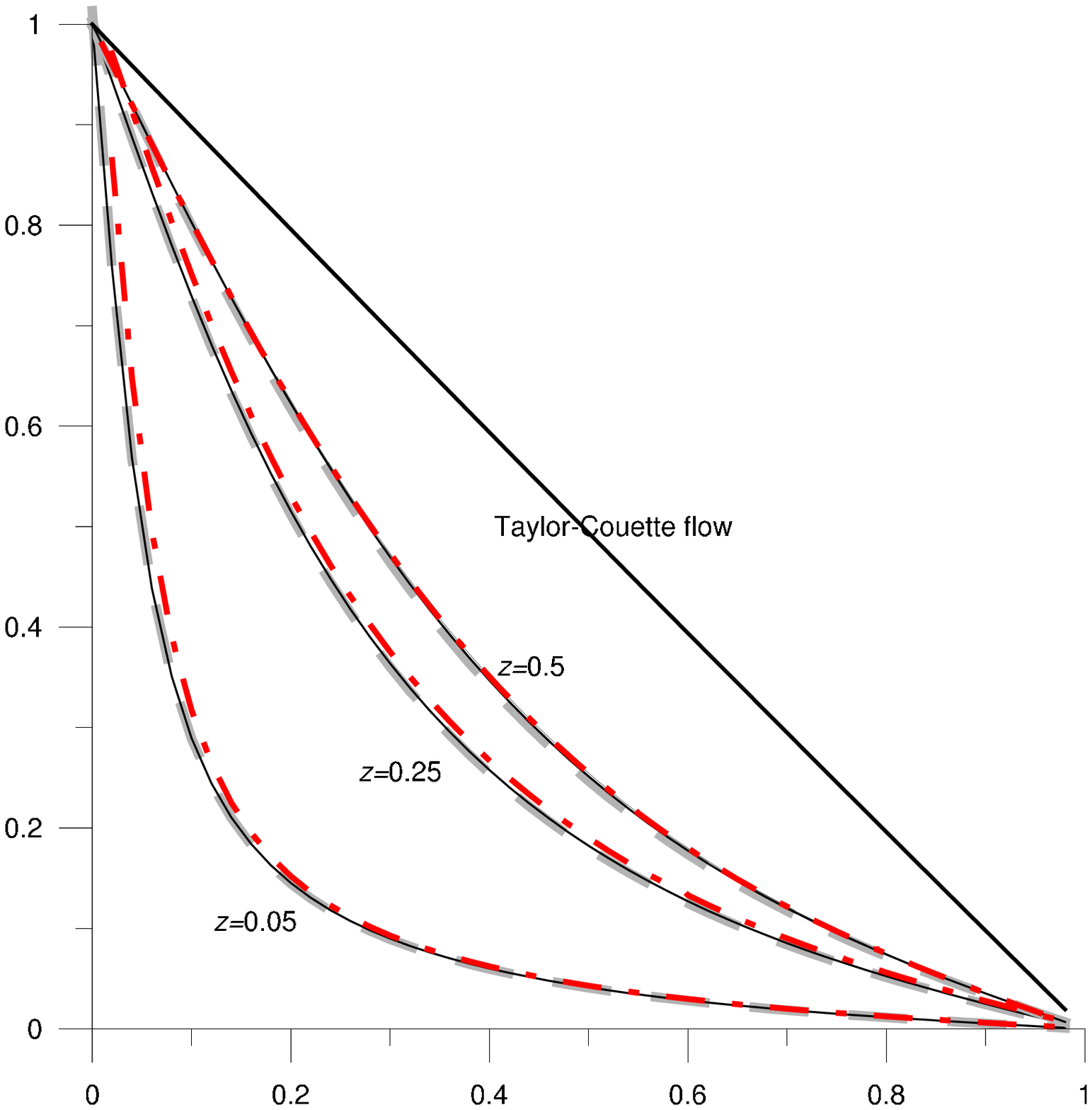}
\center a).  $R=20,$
\end{minipage} \hfill
\begin{minipage}[t]{7cm}
\includegraphics[width=7cm]{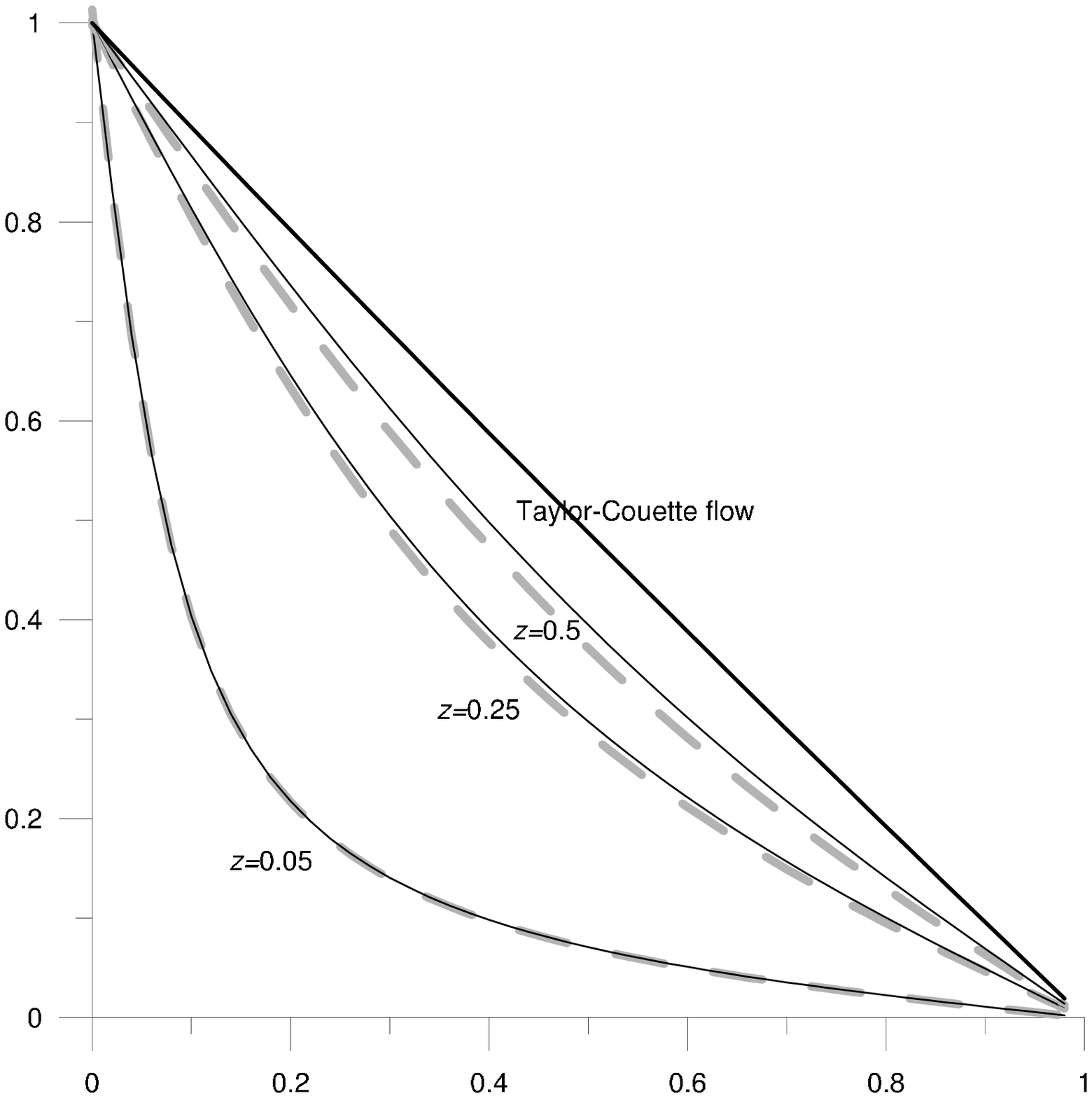}
\center b). $R=10,$
\end{minipage}
\caption{\label{prof2} The profiles $V_0(r,z)$ of the exact (dashed line), the
  approximated solutions (continous line) 
and numerical solution (dash-dotted line) on $z-$levels equal 
to $z=0.05,0.25,0.5$
  with $\Gamma=1$ as functions of $r.$
The numerical solution was found with program FLUENT
  5.5/6.0 in axisymmetric case for $\mbox{Re}=100.$ }
\end{figure}

The agreement between the approximated and the exact solutions is very
good as can be seen from fig.~\ref{prof2}.
The approximated solution (\ref{appsov}) is stable for small Reynolds
numbers $Re$ and can be used as a base state in a stability analysis.
 
We note an earlier attempt by Vladimirov \cite{Vlad} to find an
approximative solution which takes the influence of the caps into
account , which, however, differs much more from the exact solution
than ours.

\section{Stability problem for the Taylor - Couette flow under small
  perturbations}\label{loc}

\subsection{Local stability study for a short cylinder}\label{gsubs}
We consider the axisymmetrical stability problem of the Taylor-Couette
flow.  We assume that the velocity vector $\mathbf{\tilde{u}}(r,z)$
which describes the fluid motion in the annular gap like on the
fig.~\ref{k2} can be represented as a sum of two vectors
$\mathbf{u}(r,z)=(u,v,w)$ and $\mathbf{{v}}(r,z)=(0,V_{0},0)$
\begin{equation}
\mathbf{\tilde{u}}(r,z)=\mathbf{{v}}(r,z) +\mathbf{u}(r,z)=
(0,V_{0},0)+(u,v,w),
\end{equation}
where $V_{0}$ describes the fundamental mode of the Couette flow and
$\mathbf{{u}}(r,z)$ a perturbation of this mode. We assume also that
the pressure $\tilde P$ can be also represented as sum $\tilde P=P_0
+p.$

Under the following assumptions
\begin{equation}
\max_{r,z}|\mathbf{{v}}(r,z)|\gg\max_{r,z}|\mathbf{u}(r,z)|,~~\max_{r,z}%
{\frac{|\nabla\mathbf{u}|}{|\mathbf{u}|}}\gg\max_{r,z}{\frac{|\nabla
\mathbf{{v}}|}{|\mathbf{{v}}|}},\label{assumplin}
\end{equation}
where $\nabla$ denotes the spatial gradient of the corresponding
velocity, the system of the Navier-Stokes equations can be
linearized. The linearized Navier-Stokes equations (LNSE) in the
axisymmetrical case take the form
\begin{align}
\partial_{t}u-2\operatorname{Re}\frac{vV_{0}}{R+r}   =-\partial_{r}p+\Delta
u-\frac{u}{\left(  R+r\right)  ^{2}}, \label{lnse}\\
\partial_{t}v+\operatorname{Re}\left(  \partial_{r}V_{0}u+\frac{V_{0}}%
{R+r}u+\partial_{z}V_{0}w\right)     =\Delta v-\frac{v}{\left(  R+r\right)
^{2}},\nonumber\\
\partial_{t}w   =-\frac{1}{\Gamma^{2}}\partial_{z}p+\nu\Delta w,\nonumber\\
\partial_{r}u+\frac{u}{R+r}+\partial_{z}w   =0,\nonumber
\end{align}
where  $\Delta $ is the Laplace operator,
\begin{equation}
\Delta =\partial_{rr}+\frac{\partial_{r}}{R+r}+\frac{\partial_{zz}}%
{\Gamma^{2}} .\end{equation} The Reynolds number is
$\operatorname{Re}=\frac{\omega r_{in}d}{\nu} $ and $\nu$ denotes the
viscosity of the fluid.  The dimensionless variables $r,z$ describing
the flow in the cylindrical annulus belong to the intervals
\begin{equation}
r\in\left[0,1\right], ~~\ z\in\left[  0, 1 \right]. \label{bonrz}
\end{equation}

Here we use the dimensionless variables as introduced in the section
\ref{main} and additionally we use the value $d^{2}/\nu$ and $R
\omega$ to render time and velocities dimensionless too.  We leave the
same notations for the velocities in the dimensionless case as
above. It means that now $|V_{0}|\sim 1 $ and from the conditions
(\ref{assumplin}) $|v| \ll 1 $ follows.

Let's introduce a stream function $\Psi$ as usual in the axisymmetrical case
\cite{mith}
\begin{equation}
u=\frac{\partial_{z}\Psi}{R+r},
~~w=-\frac{\partial_{r}\Psi}{R+r}. \label{streamf}
\end{equation}
In terms of the stream function $\Psi$ and the azimuthal velocity $v$
the LNSE (\ref{lnse}) turns into a system of two coupled equations
\begin{eqnarray}
\partial_{t}\widetilde{\Delta}\Psi-\frac{\operatorname{Re}}{\Gamma^2}\left(  \frac
{2\partial_{z}V_{0}}{R+r}v+\frac{2V_{0}}{R+r}\partial_{z}v\right)
&=&\Delta\widetilde{\Delta}\Psi-\frac{\widetilde{\Delta}\Psi}{\left(
R+r\right)  ^{2}}, \nonumber\\
\partial_{t}v+\operatorname{Re}\left(  \left(  \partial_{r}V_{0}+\frac{V_{0}%
}{R+r}\right)  \frac{\partial_{z}\Psi}{R+r}-\partial_{z}V_{0}\frac
{\partial_{r}\Psi}{R+r}\right) &=&\Delta v-\frac{v}{\left(  R+r\right)
^{2}}.\label{azim}
\end{eqnarray}

We remark that the value $\widetilde{\Delta}\Psi$ is an azimuthal
vorticity ${\Upsilon}_{\theta}$ defined by
\begin{equation}
{\Upsilon}_{\theta}=\widetilde{\Delta}
\Psi=\frac{1}{\Gamma^2}\partial_{z}u-\partial_{r}w, ~~{\rm where}
~~\widetilde{\Delta}=\frac{\Delta }{R+r}.
\end{equation}

The basic Taylor-Couette flow is an axial inhomogeneous flow in the
case of the short cylinder because of the cylinder caps influence. The
inhomogeneity leads to additional terms which make a difference
between the system (\ref{lnse}) and the corresponding linearized
Navier-Stokes equations for an infinitely long cylinder.  Now we will
prove that the additional terms causes the propagation of dispersive
waves in the Taylor - Couette flow.

We remember that in the case of a short cylinder with very narrow gap
we have
\begin{equation}
\Gamma \sim 1,~~\eta \sim 1, ~~R \gg 1. \label{ger}
\end{equation}
Under these assumptions we can simplify the system of equations
(\ref{azim}) by replacing the coefficients $\frac{1}{(R+r)^{n}}$ (note
$r\in [0,1]$) by the constants $\frac{1}{R^n},~ n=1,2.$ The operators
$\partial _z$ and $\partial _r$ can be replaced by the multiplicative
operators $ \partial_r \sim 1/d,~~\partial_z \sim 1/l $ or, in the
dimensionless case, by
\begin{equation}
\frac{\partial}{\partial r} \sim 1,~~\frac{\partial}{\partial z} \sim
\frac{1}{\Gamma} \label{part}
\end{equation}
because of the relations (\ref{bonrz}) and (\ref{ger}). Furthermore we
assume
\begin{equation}
\frac{\partial V_{0}}{\partial r} \gg \frac{V_{0}}{R+r}
\end{equation}
and neglected $\frac{V_{0}}{R+r}$ in the system (\ref{azim}).  We
collect now the leading terms of the system (\ref{azim}) only and
obtain a simplified system of equations
\begin{eqnarray}
\partial_{t}\Delta\Psi-2\operatorname{Re}\partial_{z}V_{0}
v-2\operatorname{Re}V_{0}\partial_{z}v  &=&\Delta^2\Psi, \label{upr}\\
\partial_{t}v+\delta\operatorname{Re}\partial_{r}V_{0}\partial_{z}\Psi
-\delta\operatorname{Re}\partial_{z}V_{0}\partial_{r}\Psi &=&\Delta
v, \nonumber
\end{eqnarray}
where no-slip boundary conditions $u|_{S}=v|_{S}=w|_{S}=0$ on the
inner surface $S$ of the annulus are adopted.  In the system
(\ref{upr}) we denoted $\Delta = \partial_{rr}^{2}
+\frac{1}{\Gamma^{2}}\partial_{zz}^{2}$, and the constant
$\delta\equiv\frac{1}{R}$ is small, $\delta<<1$.

The standard, straightforward way to investigate the stability problem
would consist in finding the exact solution of the eigenvalue problem
corresponding to the system of equations (\ref{upr}). Then the
perturbations ${\bf u}(r,z)$ can be represented as a series over the
eigenfunctions. Alternatively, the system could be solved numerically
whereas a spatial discretization of the problem may be accomplished by
a solenoidal Galerkin scheme.  In our case these procedures may be
accomplished numerically only and this is outside the scope of the
present work.

Nevertheless we can give good qualitative characteristics of the flow
in the annulus if we use a local stability analysis.  This method was
first developed in the analysis of an inhomogeneous complex
astrophysical flow \cite{Gold},\cite{lifas}.  It give us the
possibility to establish the main physical processes of the considered
fluid flow system \cite{lifsh}, \cite{lifsh2}.\\[2pt]

The main idea of the local stability analysis method also named a
method of infinitesimal perturbations can be described as follows. We
suppose that the amplitude and the gradients of the main state (the
fundamental mode of the Couette flow in our case) is preserved under
the perturbations. This means we keep the values $V_0, \partial _z
V_0$ and $\partial _r V_0$ unaltered or assume that they are
constant. For all components of the perturbation vector ${\bf
u}(r,z)=(u,v,w)$ and their derivatives Taylor series are used which
are truncated after the first two terms. From the physical point of
view we assume that the perturbation scale is much smaller than the
typical scale of the axial inhomogeneity in the Taylor-Couette
system. This is the locality condition. As usual we represent the
perturbation as a plain wave,
\begin{equation}
\Psi \sim\widehat{\Psi}\exp\left(  \gamma t+i\left(
k_{r}r+k_{z}z\right)  \right) ,~~v\sim\widehat{v}\exp\left(
\gamma t+i\left(  k_{r}r+k_{z}z\right)  \right) \label{planew} ,
\end{equation}
where $\widehat{\Psi}$ and $\widehat{v}$ are the amplitudes of the
waves. The vector of wave numbers $k=(k_r,k_z)$ defines the spatial
direction of the plane wave propagation. The locality condition
requires that $k_r, k_z \gg 1.$ In (\ref{planew}), $\gamma$ has the
dimension of a frequency and in the stability analysis it is called
the time increment (or simply increment).  In the case of an intrinsic
instability of the investigated system $\gamma$ will be positive. This
means that the amplitude of the plane wave will infinitely increase
with time. This is the unstable case.

The substitution of the plane wave (\ref{planew}) into the linearized
 system (\ref{upr}) results in a system of algebraic equations for the
 increment $\gamma$,
\begin{eqnarray}
-k^{2}\left(\gamma+k^{2}\right)
\widehat{\Psi}-2\frac{\operatorname{Re}}{\Gamma^2}
\left(\partial_{z}V_{0}+iV_{0}k_{z}\right)  \widehat{v}&=&0,\nonumber\\
\left(  \gamma+k^{2}\right)
\widehat{v}+i\delta\operatorname{Re}\left(
\partial_{r}V_{0}k_{z}-\partial_{z}V_{0}k_{r}\right)  \widehat{\Psi}&=&0,\nonumber
\end{eqnarray}
where $k^{2}=k_{r}^{2}+\Gamma^{-2} k_{z}^{2}.$ This system can be
solved explicitly and we obtain two solutions for the increment
$\gamma$,
\begin{align}
\gamma_{\pm}   =-k^{2}\pm\frac{\operatorname{Re}\left(
2\delta\right) ^{1/2}\left\vert
k_{z}\partial_{r}V_{0}-k_{r}\partial_{z}V_{0}\right\vert
^{1/2}}{\Gamma k}\left(  k_{z}^{2}V_{0}^{2}+\left(
\partial_{z}V_{0}\right)
^{2}\right)  ^{1/4}\times \nonumber\\
 \times\left(  \cos\left(  \frac{1}{2}\arctan\left(  \frac{\partial_{z}%
V_{0}}{k_{z}V_{0}}\right)  \right)  -i\sin\left(  \frac{1}{2}\arctan\left(
\frac{\partial_{z}V_{0}}{k_{z}V_{0}}\right)  \right)  \right). \label{incr}
\end{align}
As a consequence, in the system two different regimes of motion can
exist.

\begin{figure}[ht]
\begin{minipage}[t]{6.8cm}
\center
\includegraphics[width=6.8cm]{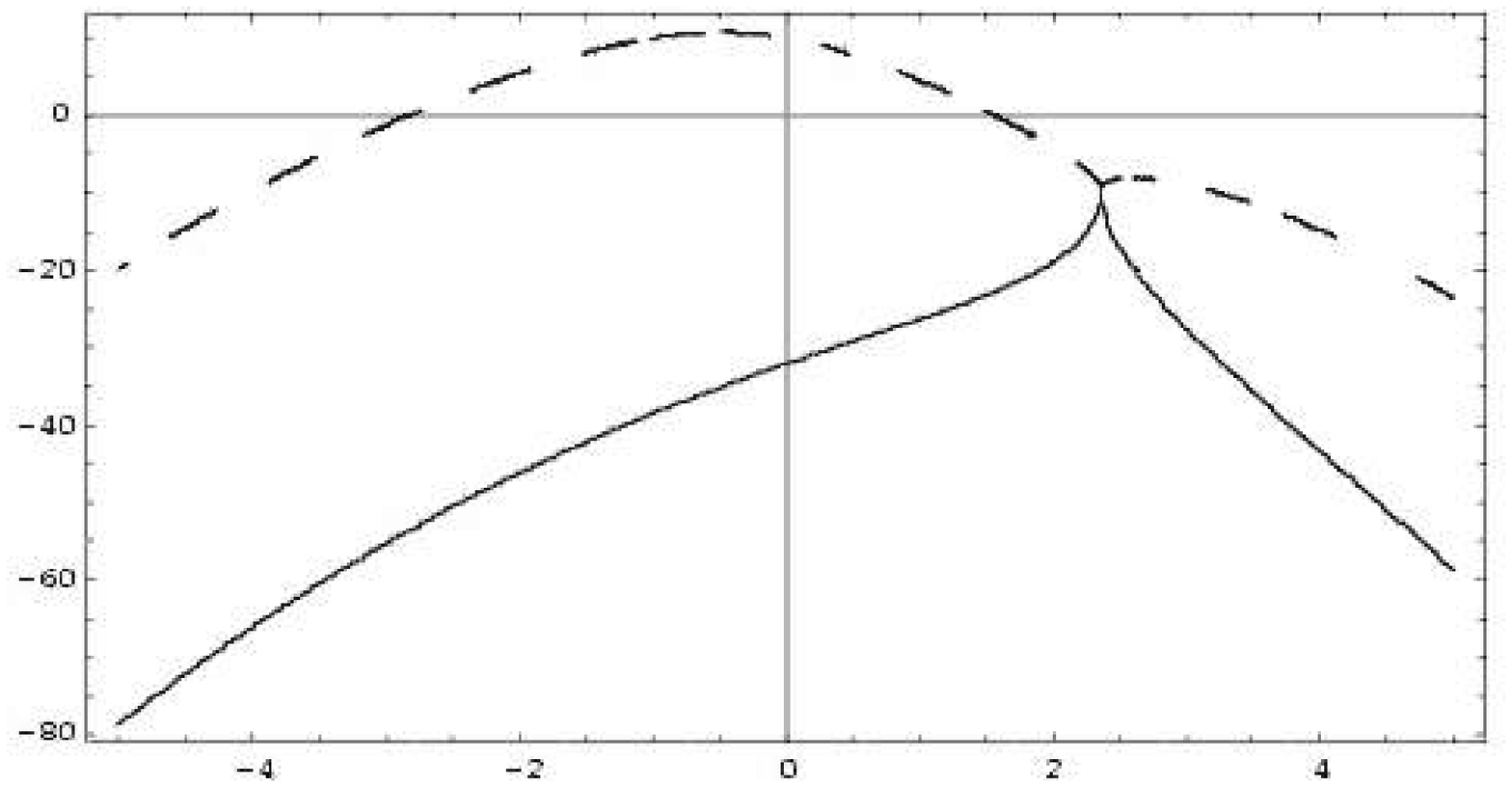}
\caption{\label{inkre} \small{The real part $Re \gamma(k_r)$ of the increment
$\gamma$ as a function of $k_r$ for $z=0.25,~ $ $r=0.50, $ $k_z=-2.00,$ 
 $\hbox{Re}=110.$}}
\end{minipage}\hspace{0.4cm}
\begin{minipage}[t]{6.8cm}
\center
\includegraphics[width=6.8cm]{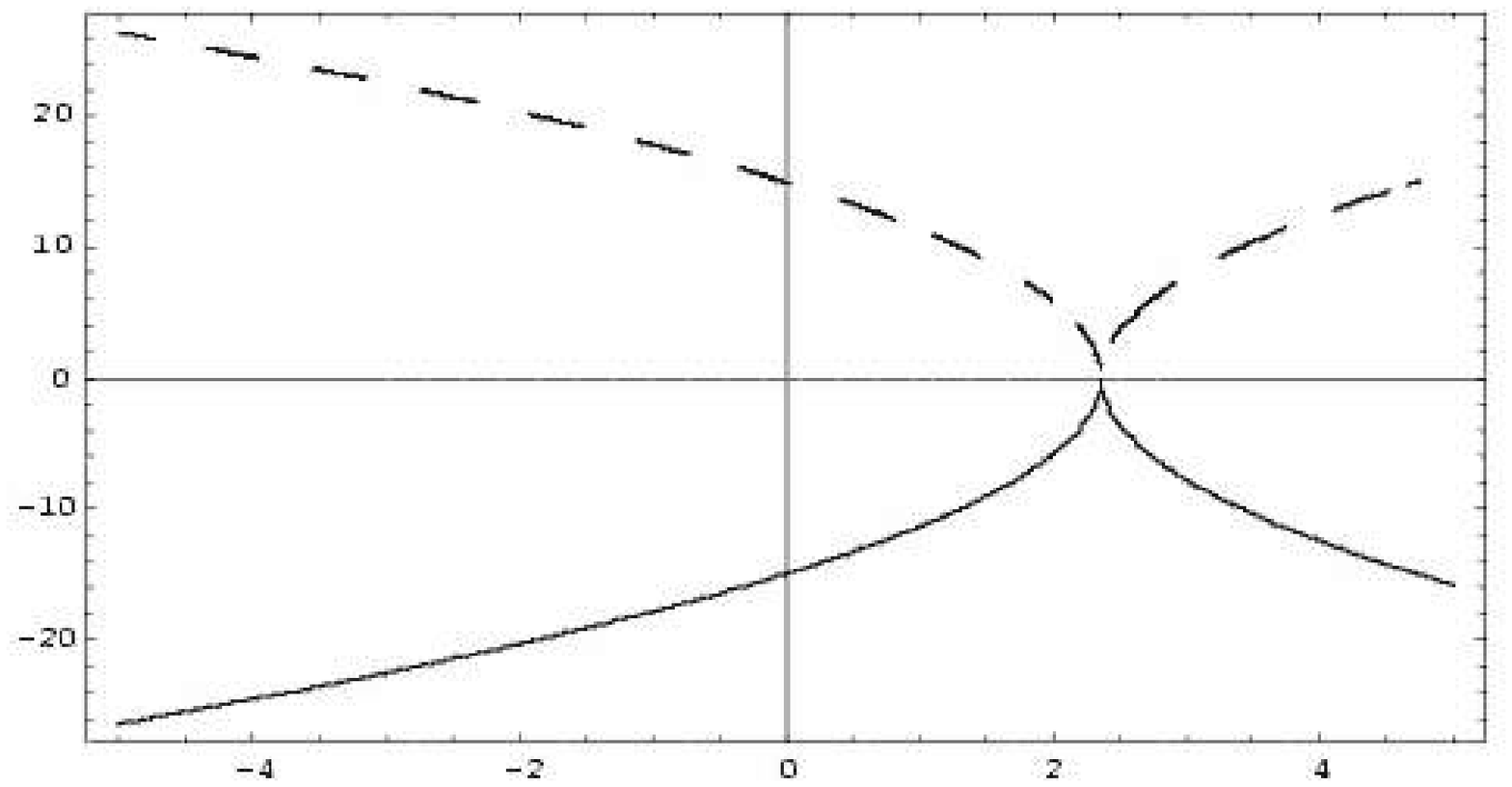}
\caption{\label{inkim} \small{The imaginary part $Im \gamma(k_r)$ of the increment
$\gamma $  as a function of $k_r$ for $z=0.25,~$ $r=0.50,$ $ k_z=-2.00,$
 $\hbox{Re}=110.$}}
\end{minipage} \hfill
\vspace{0.5cm}

\begin{minipage}[t]{6.8cm}
\center
\includegraphics[width=6.8cm]{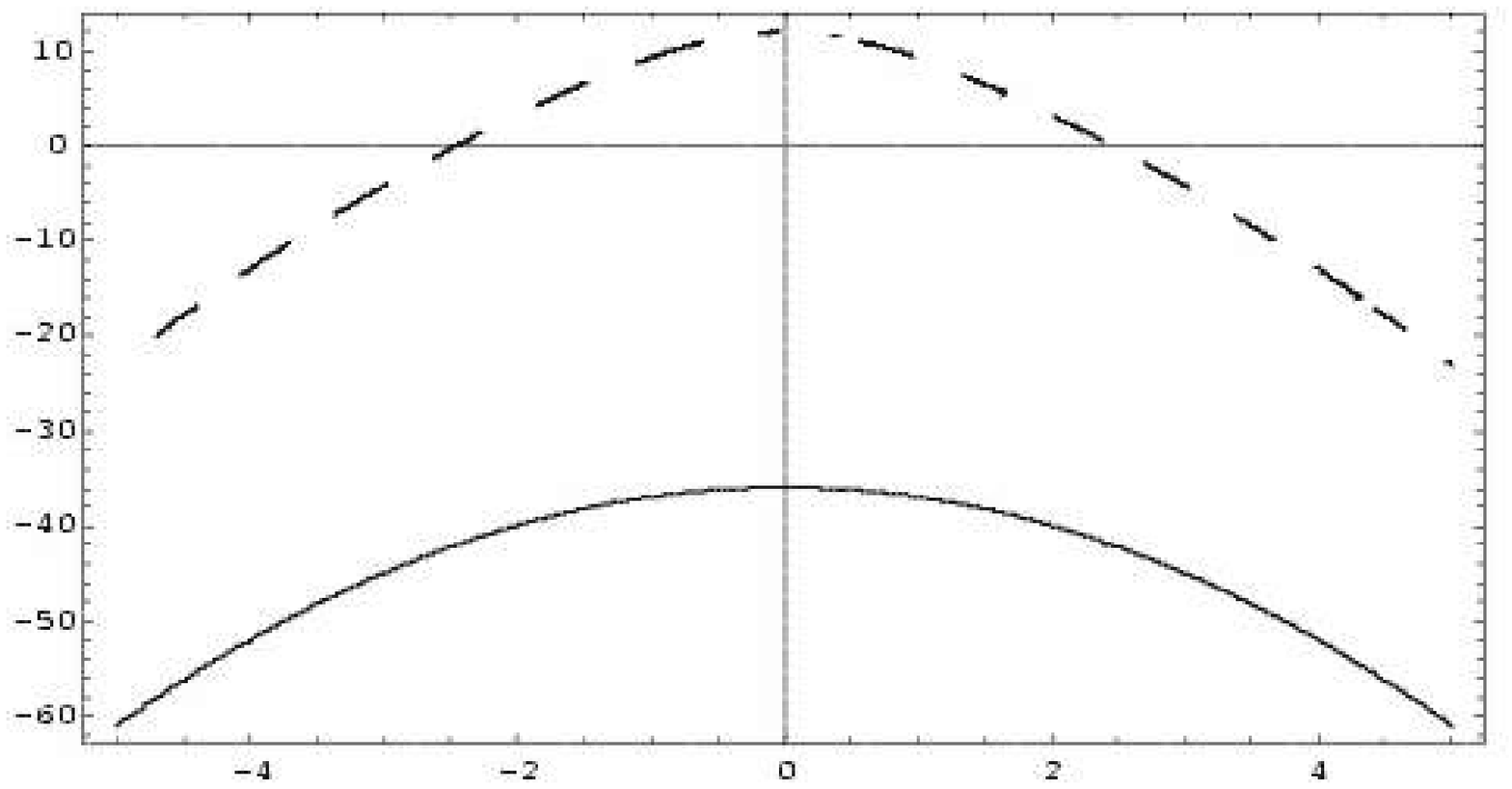}
\caption{\label{inkre1} \small{The real part $ Re \gamma(k_r) $ of the increment
$\gamma $ as a function of $k_r$ for $ z=0.50,~ $ $ r=0.5,~ $ $k_z=-2.0,$
 $ \hbox{Re}=110.$}}
\end{minipage}
\end{figure}

The motion in the annulus is strong inhomogeneous in $z-$direction
because of the influence of the cylinder caps. Let us now look for
local increments on the middle surface $r=0.5$ on three different
lines $z=0.25,~0.5,~0.75~.$ For $z=0.5$ the increment does not have any
imaginary part and on this line the system does not show any
oscillations. In this case the real part of the increment is
represented in fig.~\ref{inkre1}.  On the lines $z=0.25,~0.75$ the
increment has an imaginary part and the system has oscillations with
frequencies which are directly proportional to the gradient of the
undisturbed state.

For $ \partial_{z}V_{0}=0 $ we have two real solutions for the
increment and, consequently, two branches of non-oscillating
solutions.\\ The two branches of the increment given by
Eq.(\ref{incr}) may have also an oscillating character. For
$\partial_{z}V_{0}\neq0 $ we deal with two oscillating solutions
corresponding to waves of the inertial type. If additionally
$k_{z}\partial_{r}V_{0}-k_{r}\partial_{z}V_{0}=0 $ holds than these
two oscillating branches have an intersection point. In the
intersection point $\gamma_{\pm}=-k^{2} $ we have to do with a time
decreasing solution. On the other hand the perturbations with $
k_{z}\partial_{r}V_{0}-k_{r}\partial_{z}V_{0}\approx0 $ have the
possibility to exchange  energy.

\subsection{Local stability of the Couette flow under a small axial
  perturbation flow}
The study of the stability problem of the Couette flow was done up to
now for arbitrary small perturbations. Let's consider the effects of
an axial flow near the surface of the inner cylinder. This flow can be
generated by the pair of Taylor vortices or an axial flow through the
narrow slits in the caps of the cylinders near the surface of the
inner cylinder as shown in figure~\ref{k2}.

In order to simplify the study we suggest that the axial velocity $
w(r,z) $ has a nonzero value $ w(R,z) $ on the surface of the inner
cylinder. Additionally we suggest that the boundary condition is
homogeneous in $ z-$direction, i.e. $ w(R,z)=\zeta=const $ holds.
Here we can neglect the dependence on the $ z- $coordinate. The
constant constituent of the axial velocity distribution can be removed
by an Galilean transformation $ z^{\prime}=z-\zeta\left(1+R\right) t
;~ z^{\prime}\rightarrow z .$ The boundary condition $ w(R,z)=\zeta \ne
0 $ provides in the narrow gap approximation the linear velocity
profile in the $ r- $direction
\begin{equation}
w\approx\zeta\left(  1-\left(  r-R\right)  \right)  =\zeta\left(  1+R\right)
-\zeta r.
\end{equation}
Now we can rewrite the LNSE (\ref{lnse}) in the form
\small{ \begin{eqnarray} 
\partial_{t}\widetilde{\Delta}\Psi-\zeta\operatorname{Re}r\partial
_{z}\widetilde{\Delta}\Psi+\frac{\zeta\operatorname{Re}\partial_{z}\Psi
}{\left(  R+r\right)
^{2}}-\frac{2\operatorname{Re}\partial_{z}V_{0}}{\Gamma^2\left(
R+r\right) }v-\frac{2\operatorname{Re}V_{0}}{\Gamma^2\left(
R+r\right) }
\partial_{z}v  &=&\Delta\widetilde{\Delta}\Psi-\frac{\widetilde{\Delta}\Psi
}{\left(  R+r\right)  ^{2}}, \nonumber\\
\partial_{t}v-\zeta\operatorname{Re}r\partial_{z}v+\operatorname{Re}\left(
\partial_{r}V_{0}+\frac{V_{0}}{R+r}\right)  \frac{\partial_{z}\Psi}
{R+r}-\operatorname{Re}\partial_{z}V_{0}\frac{\partial_{r}\Psi}{R+r}
&=&\Delta v-\frac{v}{\left(  R+r\right)  ^{2}}. \nonumber
\end{eqnarray}}
We will simplify this system of equations. We remind that we
investigate the influence of small perturbations and from the
conditions (\ref{assumplin}) $\zeta\ll V_{0} $ or $ \zeta\ll1$ follows
immediately. We expand the coefficients in the system into a series
with respect to the parameter $\delta=1/R$ and keep the main terms
only.  The approximated system for the stream function $\Psi$ and for
the azimuthal velocity $v$ has the structure
\begin{eqnarray}
\partial_{t}\Delta\Psi-\zeta\operatorname{Re}r\partial_{z}\Delta
\Psi+\zeta\delta\operatorname{Re}\partial_{z}\Psi-\frac{2\operatorname{Re}}{\Gamma^2}
\partial_{z}V_{0}v-\frac{2\operatorname{Re}}{\Gamma^2}V_{0}\partial_{z}v
&=&\Delta^2
\Psi,\nonumber\\
\partial_{t}v-\zeta\operatorname{Re}r\partial_{z}v+\delta\operatorname{Re}
\partial_{r}V_{0}\partial_{z}\Psi-\delta\operatorname{Re}\partial_{z}
V_{0}\partial_{r}\Psi  &=&\Delta v. \label{axisys2}
\end{eqnarray}

As in the general case in the previous point \ref{gsubs} we will study
the stability problem under small local perturbations.  So the
perturbation scale is much smaller than a typical scale of the
inhomogeneity in the system. This locality condition implies to the
inequalities
$$k_{r}\gg1, k_{z} \gg1. $$ In the same way we assume that the values
$V_0, \partial _z V_0$ and $\partial _r V_0$ change very slowly in
comparison to the perturbations so that they can be represented by
Taylor series truncated after the first two terms.  As usual we assume
that the perturbation $\mathbf{u}(r,z)=(u,v,w)$ has the form of a wave
with time dependent amplitudes and with a time dependent wave number in
$r-$direction,
\begin{eqnarray}
\Psi &=&\widehat{\Psi}\left(  t\right)  \exp\left(  i\left(  k_{r}\left(
t\right)  r+k_{z}z\right)  \right) \nonumber ,\\
v &=&\widehat{v}\left(  t\right)  \exp\left(  i\left(  k_{r}\left(
t\right)  r+k_{z}z\right)  \right), \label{planet2}
\end{eqnarray}
where $\Psi$ is the stream function (\ref{streamf}) and $v$ is the
azimuthal velocity.  The substitution of the formulae (\ref{planet2})
into the system of equations (\ref{axisys2}) leads to the simplified
system of equations
\begin{gather}
-\partial_{t}\left(  \left(  k_{r}^{2}\left(  t\right)
+k_{z}^{2}/\Gamma^2\right) \widehat{\Psi}\left(  t\right)  \right)
-i\left(  \partial_{t}k_{r}\left( t\right)
-\zeta\operatorname{Re}k_{z}\right)  r\left(  k_{r}^{2}\left(
t\right)  +k_{z}^{2}/\Gamma^2\right)  \widehat{\Psi}\left(  t\right) +\nonumber\\
i\zeta\delta\operatorname{Re}k_{z}\widehat{\Psi}\left(  t\right)
-\frac{2\operatorname{Re}}{\Gamma^2}\left(
\partial_{z}V_{0}+iV_{0}k_{z}\right) \widehat {v}\left(  t\right)
= \left( k_{r}^{2}\left(  t\right) +k_{z}^{2}/\Gamma^{2}\right)^2
\widehat
{\Psi}\left(  t\right), \nonumber \\
\partial_{t}\widehat{v}\left(  t\right)  +i\left(  \partial_{t}k_{r}\left(
t\right)  -\zeta\operatorname{Re}k_{z}\right)  r\widehat{v}\left(  t\right)
+i\delta\operatorname{Re}\left(  \partial_{r}V_{0}k_{z}-\partial_{z}V_{0}%
k_{r}\right)  \widehat{\Psi}\left(  t\right)=  \nonumber\\
-\left(  k_{r}^{2}\left(  t\right)  +k_{z}^{2}/\Gamma^{2}\right)  \widehat
{v}\left(  t\right). \label{timedep}
\end{gather}
The solution of this system can be found for all $r$ if the 
condition  
\begin{equation}
\partial_{t}k_{r}\left(  t\right)  -\zeta\operatorname{Re}k_{z}=0, \label{keq}
\end{equation}
is fulfilled because only two terms in the system (\ref{timedep}) have
coefficients which depend on the variable $r.$ The equation
(\ref{keq}) can be easily solved and we obtain
\begin{equation}
k_{r}\left(  t\right)  =k_{r}\left(  0\right)  +\zeta\operatorname{Re}tk_{z}.
\end{equation}
The radial component $k_{r}$ of a wave number vector $\bf{k}$
increases or decreases linearly in dependence of the sign of the axial
component $k_{z}$ of the wave number vector.  In another words we can
explain the action of the axial shear as a change of the radial scale.

The time dependence of the wave number $k_{r}(t)$ leads to an
algebraic non-exponential evolution which can be described by the
system of equations
\begin{gather}
-\partial_{t}\left(  \left(  k_{r}^{2}\left(  t\right)
+k_{z}^{2}/\Gamma^2\right) \widehat{\Psi}\left(  t\right)  \right)
+i\zeta\delta\operatorname{Re} k_{z}\widehat{\Psi}\left(  t\right)
-\frac{2\operatorname{Re}}{\Gamma^2}\left(  \partial
_{z}V_{0}+iV_{0}k_{z}\right)  \widehat{v}\left(  t\right)=  \nonumber\\
\left(  k_{r} ^{2}\left(  t\right)
+k_{z}^{2}/\Gamma^{2}\right)^2 \widehat{\Psi}\left(
t\right), \nonumber  \\
\partial_{t}\widehat{v}\left(  t\right)  +i\delta\operatorname{Re}\left(
\partial_{r}V_{0}k_{z}-\partial_{z}V_{0}k_{r}\right)  \widehat{\Psi}\left(
t\right)  =-\left(  k_{r}^{2}\left(  t\right)
+k_{z}^{2}/\Gamma^{2}\right) \widehat{v}\left(  t\right).
\label{cylnse}
\end{gather}
Perturbations as studied here were considered in fluid dynamics since
Lord Kelvin, but it has been realized only recently that these
perturbations can play a trigger role in the subcritical turbulence
transition.  These untypically properties of the flow resulting from
the fact that the corresponding LNSE operator is non-self-adjoint for
shear flows such as a plane Couette flow \cite{Chag1}, \cite{Butf}, a
Poiseuille flow \cite{Red}, a counter-rotating Taylor-Couette flow
\cite{TTC1}, \cite{TTC2}, a boundary layer \cite{Gust}, a Kepler rotation
\cite{Chag1}. Apart from the modes of the discrete spectrum there
exist plenty of modes of the discontinuous spectrum. The algebraic
time evolution of these modes can achieve large amplitudes which can
reach hundredfold of the initial values of amplitudes. Another
remarkable property of these modes is their trait to redistribute
energy and momentum between different types of perturbations
\cite{Shear1}, \cite{Shear2}.

To give the qualitative description of an impact from an axial
flow on the stability of the Couette flow we will find the
increment for a fixed instant time. Analogously to (\ref{planew})
we have the representation 
\begin{equation}
\widehat{\Psi}\left(  t\right)\sim \exp \left(  \gamma
  t\right),~~\widehat{v}\left(  t\right)\sim \exp \left(  \gamma t\right) \label{inkax}
\end{equation}
for the time dependent amplitudes of the wave (\ref{planet2}). If we virtually
hold unaltered the time-dependence of $k_{r}\left( t\right) $ we can
consider the stability analysis local in time. The substitution of the
formula (\ref{inkax}) into the system of equations (\ref{cylnse})
leads to an algebraic equation for the increment. We get now two
solutions of this system for the increment $\gamma$
\begin{align}
\gamma_{\pm} &  =-k^{2}+\frac{i\zeta\delta\operatorname{Re}k_{z}}{2 k^{2}%
}\pm \nonumber\\
& \frac{\operatorname{Re}\delta^{1/2}\left(  64k^{4}\Lambda^{2}\left(
\partial_{z}V_{0}\right)  ^{2}+\left(  \zeta^{2}\delta k_{z}^{2}+8k^{2}%
k_{z}V_{0}\Lambda\right)  ^{2}\right) ^{1/4}}{2\Gamma
k^{2}}\exp\left(
-i\Phi\right)  ,\\
\Lambda &  =k_{z}\partial_{r}V_{0}-k_{r}\partial_{z}V_{0},~~~~ \Phi=\frac{1}%
{2}\arctan\left(  \frac{8k^{2}\partial_{z}V_{0}\Lambda}{\zeta^{2}\delta
k_{z}^{2}+8k^{2}k_{z}V_{0}\Lambda}\right).  \nonumber
\end{align}
For small $\zeta\operatorname{Re}$ the dispersion curves have the same
form as considered in the previous section (without any axial flow).
In the intersection area $k_{z}\partial_{r}V_{0}-k_{r}
\partial_{z}V_{0}\approx0$ we have  now
\begin{equation}
\gamma_{+}=-k^{2}+\frac{i\zeta\delta\operatorname{Re}k_{z}}{2 k^{2}%
}, ~~\gamma_{-}=-k^{2}.\label{zeroinkr}
\end{equation}
This dispersion equation is similar to the equation for the Rossby
waves for inhomogeneous rotation of the Earth or othetrs systems.  
The wave number $k_r(t)$ is no longer fixed as in the previous case
(\ref{planew}) but it has a time dependent value. As a consequence,
the point $Im \gamma\left(k_r(t)\right)$ or $Re
\gamma\left(k_r(t)\right)$ will move along the branches (similar to
the branches on the fig.~\ref{inkre}~-~fig.~\ref{inkre1}) with time.
Chagelishvili and Chkhetiani \cite{Shear1} have proved that under the
influence of the shear the flow frequencies of the Rossby waves and
inertial waves will be modified so that they take the same values.
This circumstance leads to the energy exchange between different modes
in the flow.  The same mechanism of the energy exchange is
characteristic for the system (\ref{cylnse}).

Suppose that we start the motion near the intersection point. One mode
is unstable and starts to grow while the second mode will
dissipate. In the intersection point their frequencies are very close
to each other and a resonance interaction may take place.  In this
case the dissipative mode loses the main part of its energy. Thus, we
have found a mechanism for the instability which may be responsible
for transport and exchange of energy and angular momentum.

\begin{figure}[ht]
\begin{minipage}{7cm}
\center
\includegraphics[height=3.8cm]{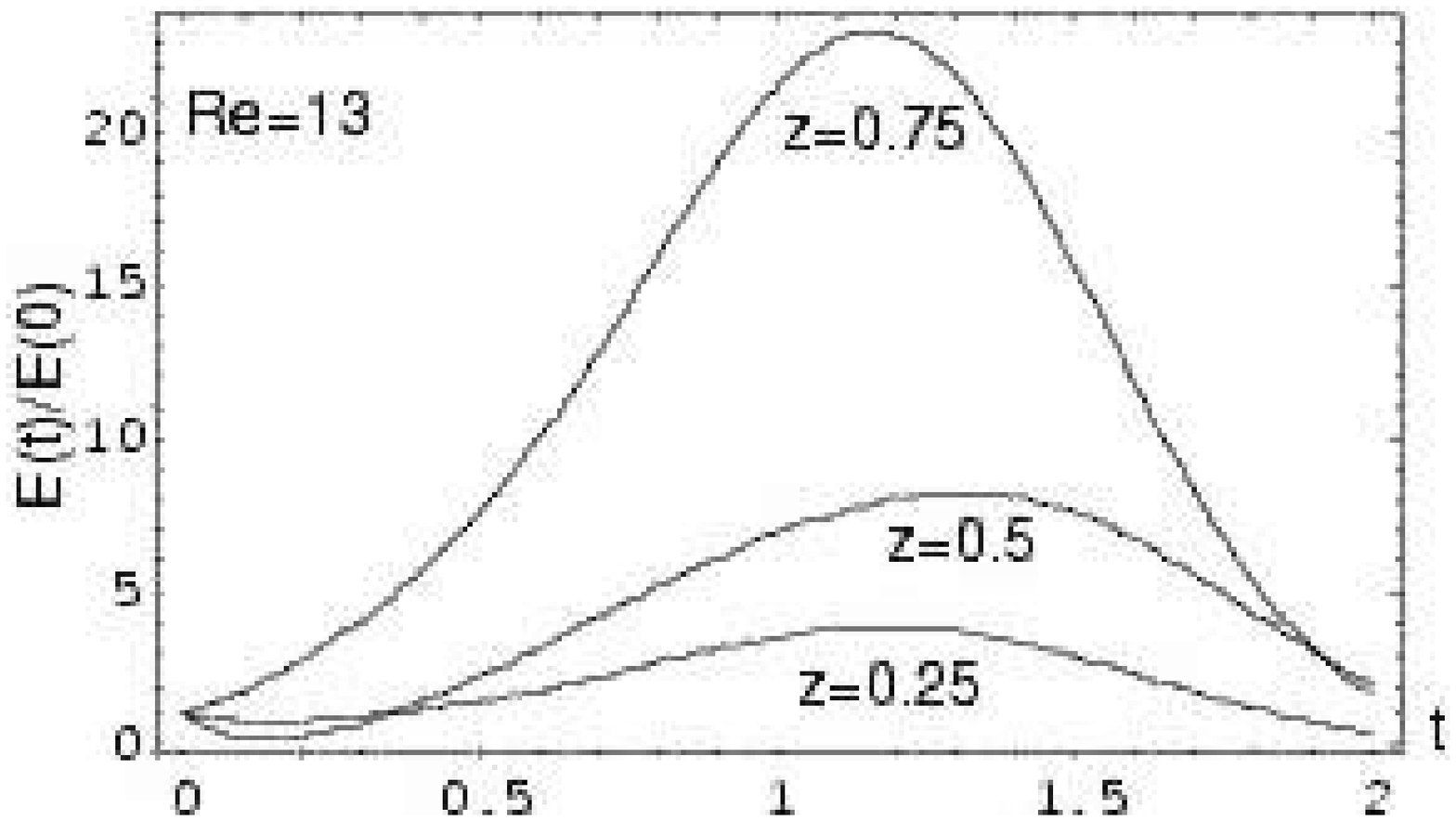}
\center a) $\hbox{Re}=13$
\end{minipage}
\begin{minipage}{7cm}
\center
\includegraphics[height=3.9cm]{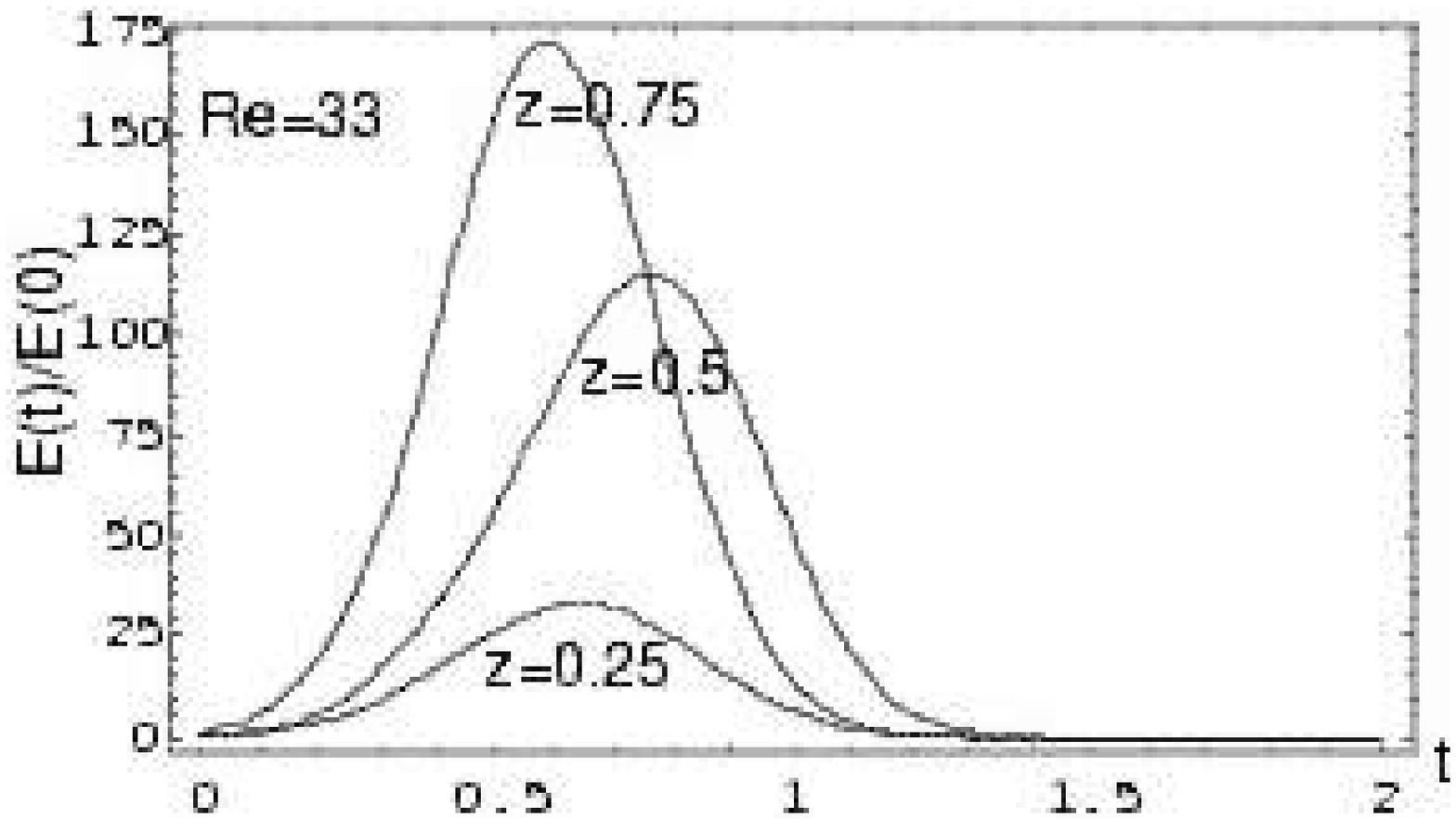}
\center b) $\hbox{Re}=33$
\end{minipage}
\caption{\label{inkim2} \small{The time dependent energy growth for
 $ \Gamma=1,$ $ \zeta=0.1~, $ $ r=0.5~,$ $ k_r(0)=1.0,$ $ k_z=-1,$
$ \widehat{\Psi}(0)=1.0, $ $ \widehat{v}(0)=1$
for different levels of $z$, $z=0.25~,~0.50~,~0.75.$}}
\end{figure}

If we analyze the solutions of the system (\ref{cylnse}) we notice
that axial flow leads to a strong differentiation in the energy growth
of the perturbations in dependence on the axial level.  This
dependence is represented for two Reynolds numbers in
fig.~\ref{inkim2}. This asymmetry can be the seed for the symmetry
breaking in the Taylor-Couette system in short cylinders.

We can assume that the same mechanism work also without any axial flow
in the system. Consider two symmetrical counter rotating Taylor
vortices in the annulus. Then we have in the annular gap a radial
distribution of the axial velocity component which is oppositely
directed in different parts of the annular gap.  If now a small
perturbation amplified the circulation in one of the vortices for a
small interval of time then radial gradients of the axial velocity
will show up in different parts of the gap. Therefore the degree of
growth of perturbations in the continuous spectrum will be different
in both parts of the gap and the faster growing perturbation can
trigger an irreversible process of symmetry breaking.

\section{A symmetry breaking and helicity conservation law}\label{helica}

The classical Taylor - Couette problem with pure rotation of one or
both cylinders posses $\bf Z^2$ and $\bf SO(3)$ symmetries. The
implementation of an axial flow immediately destroyes the $\bf Z^2$
mirror symmetry. The axial perturbation flow amplifies in-homogeneously
different intrinsic modes in the studied states as we proved in the
previous section. It results in a reach variety of super critical
stationary and quasi periodical states. This symmetry breaking for
short cylinders can be considered as a generation of certain helicity
states.

The helicity is a topological characteristic of the flow and it is the
second inviscid integral invariant for the Euler equations.  The
helicity is defined by
\begin{equation}
\mathcal{H}=\mathop{\displaystyle \int}\limits_{V}\mathbf{u}\cdot\left[
\nabla\times\mathbf{u}\right]  dV,
\end{equation}
where the vector $\Upsilon\equiv \nabla\times\mathbf{u}$ defines the
vorticity of the flow.  The helicity is a pseudo scalar because of
\[
\mathcal{H}\left(  -\mathbf{r}\right)  =-\mathcal{H}\left(  \mathbf{r}\right).
\]
It is well known that the helicity is an important quantity in the
study of dynamics and stability of complicated vortex structures and
flows.  The states with maximum of helicity correspond to the states
with minimum of energy \cite{Moffat92}, \cite{ Chkhe01}, \cite{Chkhe02}.

In the same way as we get the evolution equation for the energy of the system
\[
\partial_{t}\mathcal{E}=\mathop{\displaystyle \oint}\limits_{S}p\cdot
\mathbf{u}_{\mathbf{n}}dS-\nu\mathop{\displaystyle
\int}\limits_{V}\left[ \nabla\times\mathbf{u}\right]  ^{2}dV
\]
we obtain the following evolution equation for the helicity,
\begin{equation}
\partial_{t}\mathcal{H}=\mathop{\displaystyle \oint}\limits_{S}\left[
\nabla\times\mathbf{u}\right]  _{\mathbf{n}}\left(  \frac{\mathbf{u}^{2}}%
{2}-\frac{p}{\rho}\right)  dS-\nu\mathop{\displaystyle \int}\limits_{V}\left[
\nabla\times\mathbf{u}\right]  \cdot\left[  \nabla\times\left[  \nabla
\times\mathbf{u}\right]  \right]  dV, \label{evolhel}
\end{equation}
where $\mathbf{n}$ is the unit normal vector orthogonal to the boundary
surface $S.$

For the inviscid fluids the right hand side of equation
(\ref{evolhel}) reduces to the first term. The helicity conserves if
the normal component of the vorticity on the boundary surface
disappears, $\left[ \nabla\times\mathbf{u}\right] _{\mathbf{n}}=0$, or
if we have a potential flow.  Let $S$ include a bounded volume of an
ideal fluid with $\Upsilon \ne 0$ and $\mathcal{H} \ne 0$ then the
helicity will be conserved by the motion.  For very small viscosity
$\nu \ne 0$ the helicity is no longer conserved but it changes very
slowly because of the negligibleness of the second term in
(\ref{evolhel}). As a consequence, for large Reynolds numbers we can
treat the helicity as an approximately conserved quantity.

In a classical Taylor-Couette system with mirror symmetric boundary
conditions we expect an even number of counter-rotating vortices.  The
growth of anomalous modes and the phenomenon of appearing of
time-dependent states can be considered as a mirror symmetry breaking
process and a generation of states with a definite sign of the
helicity.  Let us represent the helicity in the cylindrical coordinate
system and reduce the formula to the case of short cylinders.  The
components of the vorticity $ \Upsilon $ are
\begin{equation}
\Upsilon_{r}=-\partial_{z}v,~~\Upsilon_{\theta}=\partial_{z}u-\partial_{r}
w,~~\Upsilon_{z}=\partial_{r}v+\frac{u}{r}. \label{vort}
\end{equation}
Using the expressions (\ref{vort}) we obtain for the helicity
$\mathcal{H}$ in the axisymmetrical case the convenient representation
\[
\mathcal{H}=4\pi\mathop{\displaystyle \iint}\limits_{S}\left(  v\partial
_{z}u+w\partial_{r}v+\frac{wv}{r}\right)  rdrdz.
\]
For a narrow annular gap with $\eta\sim1$ and $R\gg 1$ we can
approximate the formula for the helicity by
\[
\mathcal{H}\cong4\pi R\mathop{\displaystyle \iint}\limits_{S}\left(
v\Upsilon_{\theta}+\frac{wv}{r}\right) drdz\approx4\pi R 
\mathop{\displaystyle \iint}\limits_{S}v\Upsilon_{\theta}drdz.
\]
The axial flow generates a constant stream of helicity and angular
momentum through the annulus. It is an important source of the
symmetry breaking in the system.

\section{Numerical facilities}\label{sec7}

Major advances in computer technologies and numerical techniques have made
possible to propose an alternative or at least a complementary approach
to the classical and analytical techniques used in laboratory experiments.
Computational Fluid Dynamics becomes part of the design process.
But discontinuities or deep gradients lead to computational difficulties
in the classical finite difference methods, although for the finite element
methods often a
suitable variational principle is not given. Therefore, we consider the finite
volume methods, which are based on the integral form over a cell instead of
the differential equation. Rather than a pointwise approximation at the grid points,
we break the domain into \textit{grid cells} and approximate the total
integral over each grid cell. It is actually the \textit{cell average}, which
equals this integral divided by the volume of the cell.

We consider the axial symmetric case as well as the full 3D flow in
the annulus. The numerical method allows to choose a so-called
segregated or a coupled solver. The simulation process consists of
(see e.g. \cite{MF})
\begin{itemize}
\item division of the domain into discrete control volumes
  (construction of the computational grid),
\item integration of the governing equations (in axial-symmetric or in
  full 3D form) on the individual control volumes to construct
  algebraic equations for the discrete variables (velocity, pressure,
  temperature and other parameters),
\item linearization and solution of the algebraic system to yield
  updated values of the variables.
\end{itemize}
Both the numerical methods of the segregated and the coupled solvers
employ a similar finite volume process, but the approach used to
linearize and solve the equations is different.

\begin{figure}[ht]
\center
\includegraphics[width=4.8cm, angle=270]{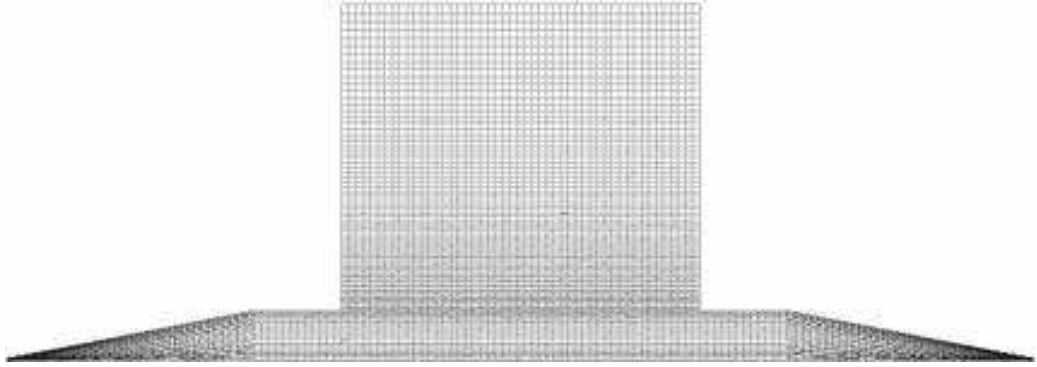}
\caption{\label{2dgrid} \small 2D structural grid for the annulus and the slits with 7690 quadrilateral cells.}
\end{figure}

The FLUENT 5.5/6.0 \cite{FLUENT} simulation and post-processing program was applied
for the numerical calculation of the flow. The unsteady 2D
axisymmetric or 3D version of the Navier-Stokes equations
is solved as described using finite volume discretization on a structured
quadrilateral grid (fig. {\ref{2dgrid}}). 
The finite volume method has a first order upwind scheme and constant
relaxation factors, while the SIMPLE procedure for the
pressure-velocity coupling is applied.

The grid is generated with the help of GAMBIT program and
characterized by a large difference (100 times) between boundary
edges. But it was possible to save the grid structural and
quadrilateral.  The smallest inflow and outflow faces of the grid are
divided uniformly by 20 nodes, whereas the nodes in the bigger
geometry of the axial flow are a more concentrated near the boundary
because some vortices can arise there if the Reynolds number is large
\cite{LAD}. The major volume of the annular gap is divided 50x70
nodes, uniformly in the axial direction and linear condensing in the
radial direction.

As next, we set up a 2D axisymmetric model, using a rotating reference
frame.  The time step is selected to be $\frac{1}{80}$ of a single
rotation.  It helps to determine the flow motion state in the quarters
of the rotation circle ($\theta=0, \pi/2,\pi, 3 \pi/2$). Each time
step includes 30 iterations or fewer if the solution converges with
the residual of order $O(10^{-5})$.

\begin{figure}[ht]
\center
\includegraphics[width=9.5cm, angle=270]{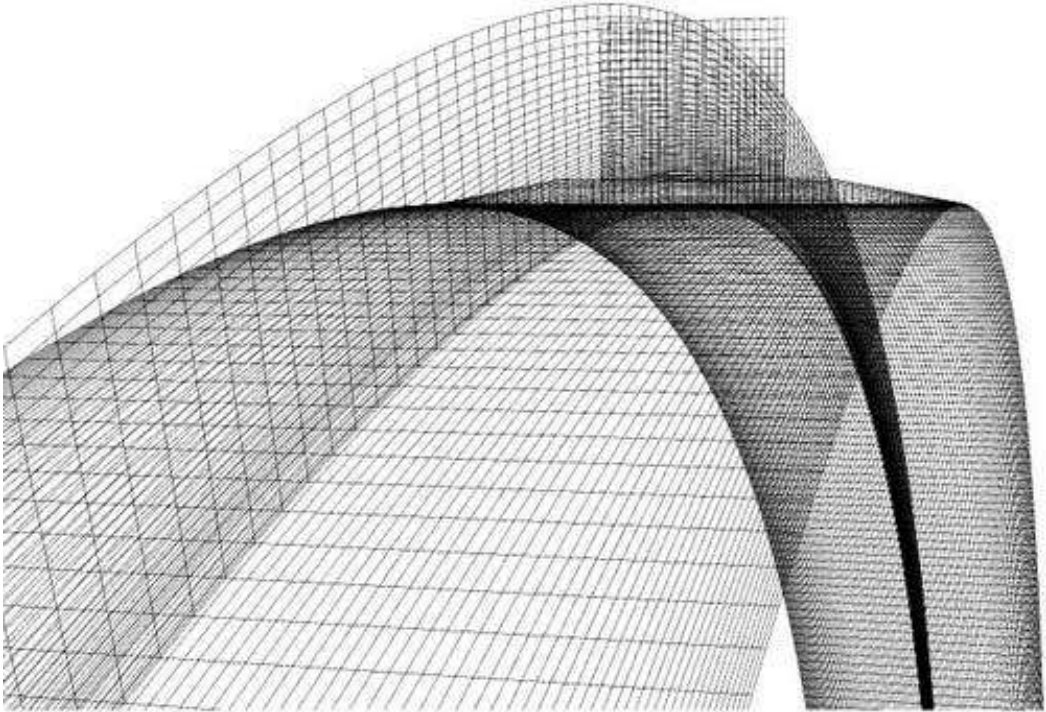}
\caption{\label{3dgrid} \small 3D grid with 1087040 hexahedral cells.}
\end{figure}

The plane of the 3D grid that corresponds to 2D axisymmetric model
have to be coarser, because of the large diameter of the inner
cylinder and therefore of the large number of cells in the projection
on the $(r,\theta)$-plane. These are only 10 nodes on the inflow and
outflow faces in the conical tapering of the slits and 25x30 in the
annulus of the slits (fig. \ref{3dgrid}).
It results in 1087040 hexahedral cells, that is close to the limit for
a computation and post-processing on a single processor in an
acceptable time. Another, more complicated 3D simulation would be
possible using parallel computing \cite{MY}. Such a program is used to
being written with the Help of Message Passing Interface (MPI)
\cite{MPI} or some other specific approaches \cite{MFR}. But this is
the aim of further investigations and developments.

Some test simulations showed, that the 3D case is not sensitive enough
to determine the very small scale instabilities correctly. Although,
the results of the 3D and 2D axisymmetric simulations on the same grid
are equivalent.  It was verified using the same coarser grid for the
2D case or by solution of the stationary problem without
instabilities.  A finer grid for the 2D axisymmetric case, by
contrast, does not give more substantial changes to the behavior of
the unstable flow. So, the 2D axisymmetric case (fig.~\ref{2dgrid})
with 7690 quadrilateral grid cells is chosen and can be considered
without loss of accuracy for all computations in that geometry.

\section{Pattern structure in the annulus under influence of the axial flow}\label{resul}

\begin{figure}[htbp]
\begin{minipage}{9.5cm}
\includegraphics[width=7.5cm, angle=270]{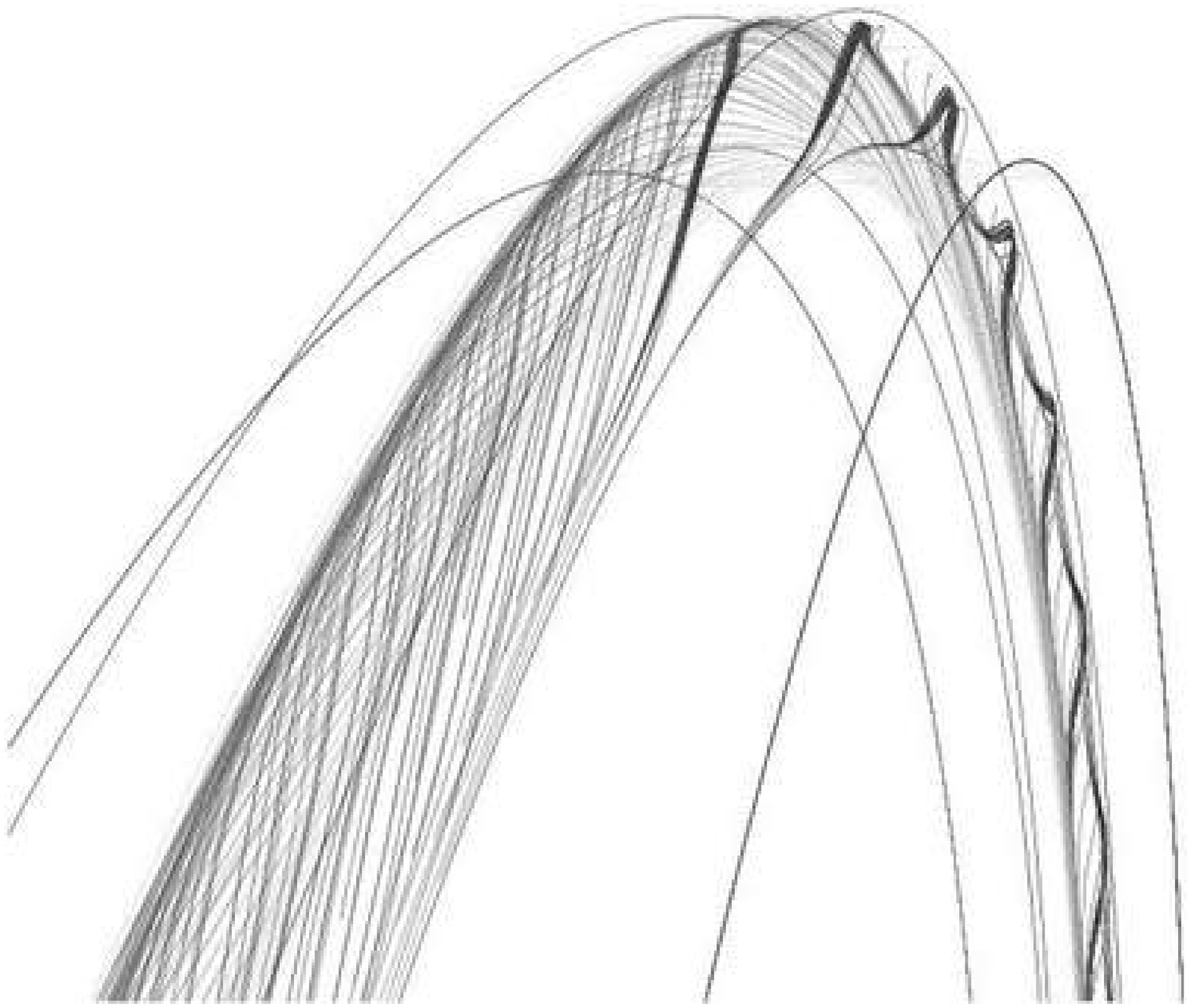}
\end{minipage}
\begin{minipage}{4.4cm}
\center
\includegraphics[width=2.6cm, angle=270]{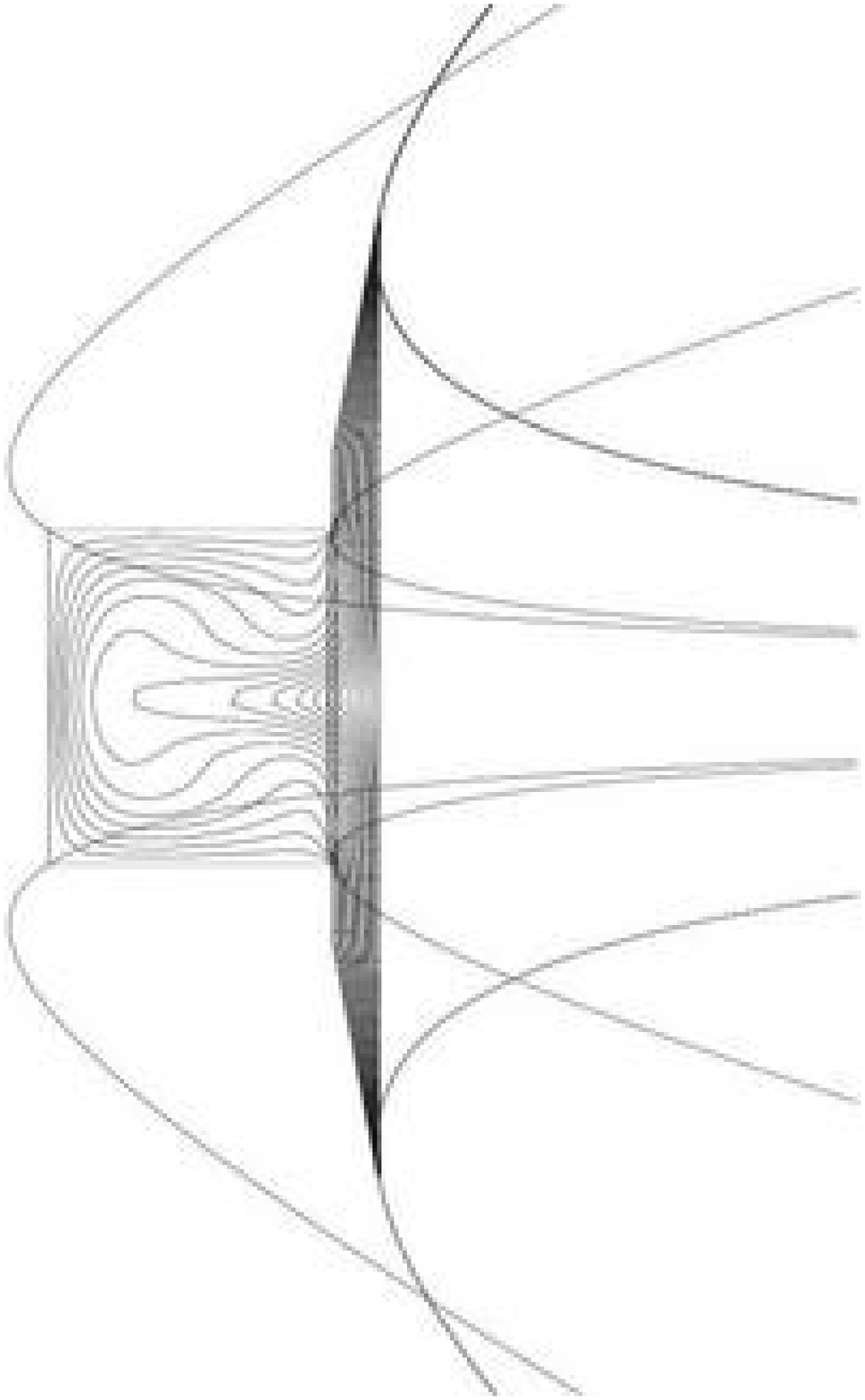}
\includegraphics[width=2.5cm, angle=270]{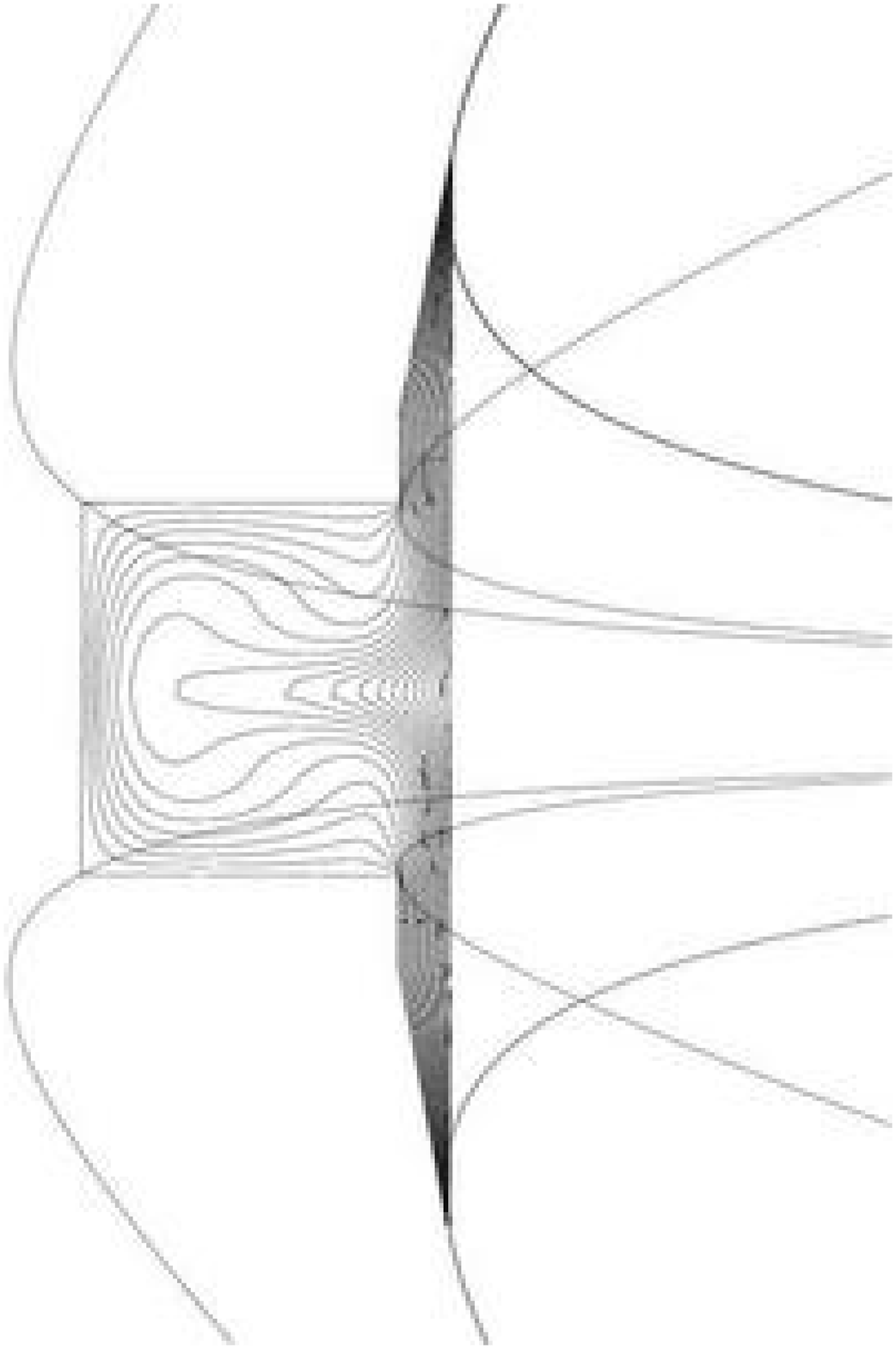}
\includegraphics[width=2.1cm, angle=270]{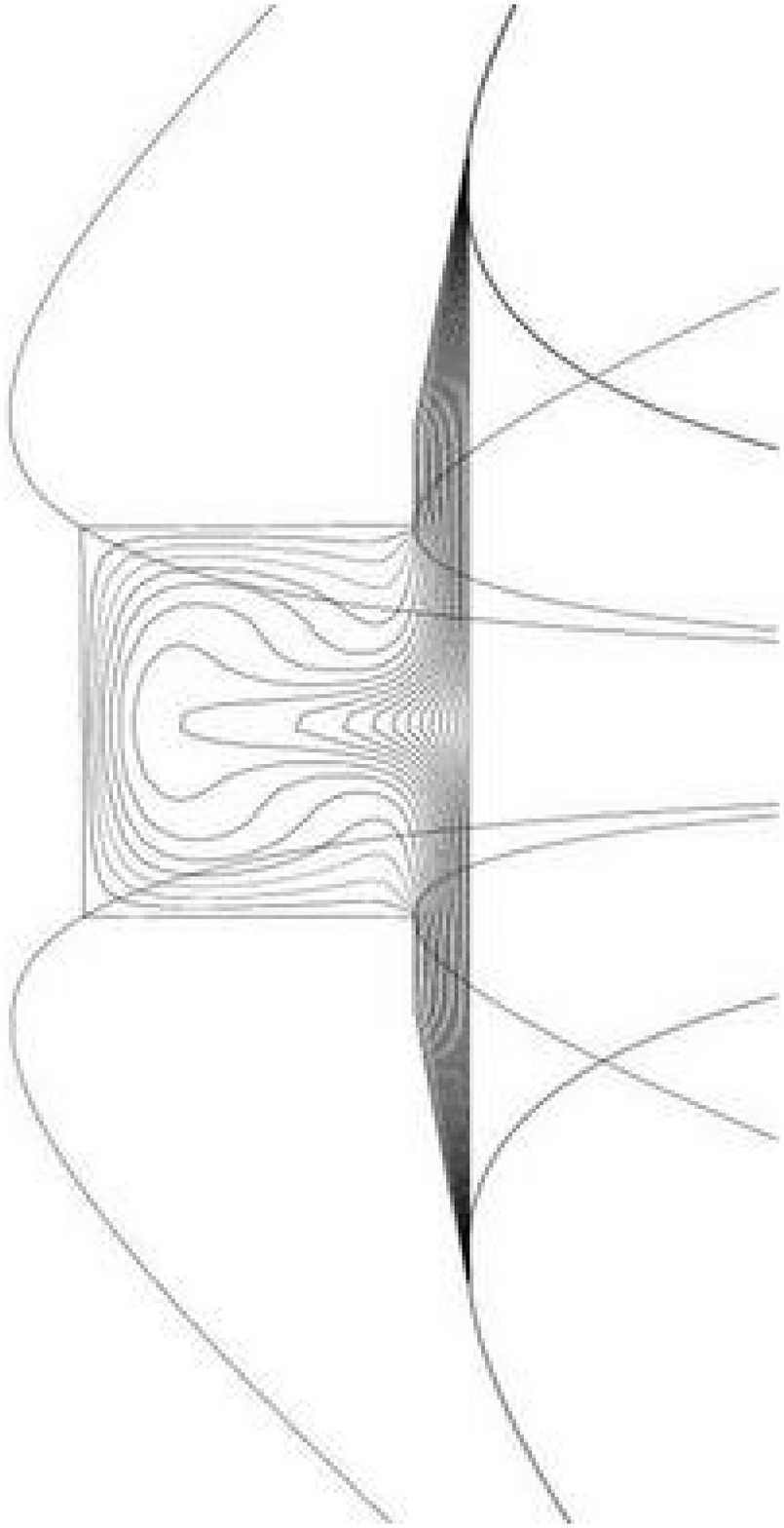}
\end{minipage} 
\caption{\label{3Dri} 3D simulation of the two symmetrical stationary
vortices.  Only one of the vortices is represented to avoid an
overload of the left figure.  The counter lines of the velocity
magnitude are given in 3D case on the left and in cross sections on
the right right by $\theta =0, \pi/2, \pi$, where $\nu=0.1 (
10^{-3}\frac{m^2 }{ s}),$ $w=0.016~,$ $u_{max}=0.060~, $ $
\mbox{Re}=1256.0 ~,$ $ \mbox{Re}_{ax}=0.1~.$}
\end{figure}

In numerical experiments which simulated the flow in the annular gap
as in fig.~\ref{k2} and more detailed in fig.~\ref{2dgrid} both axial
symmetrical and 3D cases were studied.  3D calculations had been done
prepared for a large number of examples. But it became apparent that
no torsion or other remarkable 3D effects arose in the system for
$\mbox{Re} \le 6000$. Therefor we used for further numerical
investigations axisymmetrical case which saved a lot of time and
computer capacity. As a graphical representation of the results in 3-D
cases on a paper sheet does not show the interesting region in sufficient detail so
that we represent only one fig.~\ref{3Dri} which gives an impression,
that the processes under investigation are indeed mostly
axisymmetrical.

In all investigations the inner cylinder rotates with the same
azimuthal velocity $V_0= 2 \pi \omega r_{in}$ on the surface of the
inner cylinder. We used this value to render other velocities
dimensionless. The weak axial flow is directed on all figures from
left to right.  The entrance and outgoing slits in the ends of conical
tapering have the radius ratio $\tilde{ \eta }=0.99992.$

We found that there exist four different flow states with typical
patterns in dependence on the relation between azimuthal and axial
Reynolds numbers.  The azimuthal Reynolds number $\mbox{Re}$ is
defined by $\mbox{Re}=V_0~ d / \nu$ and the axial Reynolds number by
$\mbox{Re}_{ax}=w_{max}~ d / \nu. $

\begin{figure}[htbp]
\center
\includegraphics[width=140mm]{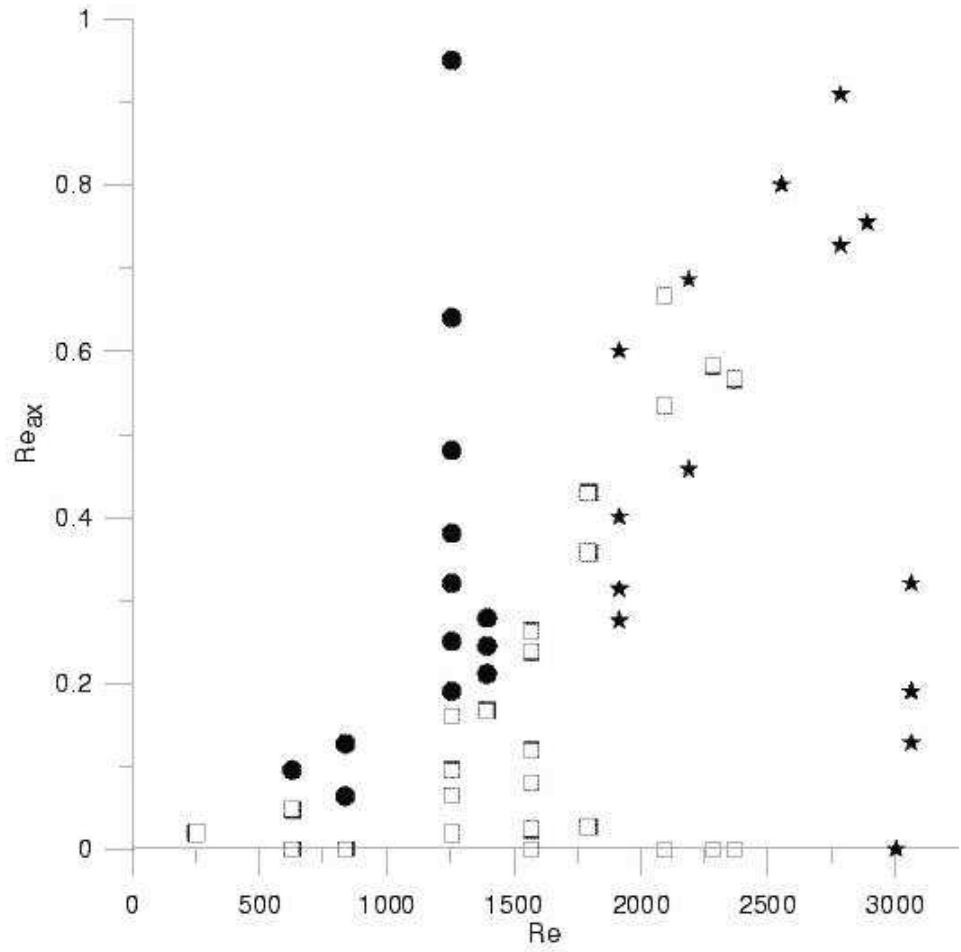}
\caption{\label{tab1} The pattern distribution in dependence on
the azimuthal $\mbox{Re}$ and axial $\mbox{Re}_{ax}$ Reynolds numbers. Two Taylors vortices
in the annulus are denoted by $\square$, one stable vortex by
{\Large$\bullet$}, natural oscillation states of two vortices by
$\bigstar .$}
\end{figure}

We studied a pure rotation and then switched on an axial flow to
investigate the sensitivity of the flow patterns to small axial
perturbations.  The pattern structure is very sensitive to the
appearance of an axial flow.  We looked for very weak axial flow with
$\mbox{Re}_{ax}\ll \mbox{Re} $ nevertheless the pattern structure
changed immediately under the influence of the axial flow.  We studied
the states distribution in the regions $\mbox{Re} \in[0,12000]$ and
$\mbox{Re}_{ax} \in[0,10].$ The results for the most interesting part
are that for $\mbox{Re} \in[0,2500]$ and $\mbox{Re}_{ax} \in[0,1]$ and
they are represented on the schematic diagram fig.~\ref{tab1}.

\begin{figure}[htbp]\unitlength1cm
\center
\includegraphics[ width=45mm,angle=270]{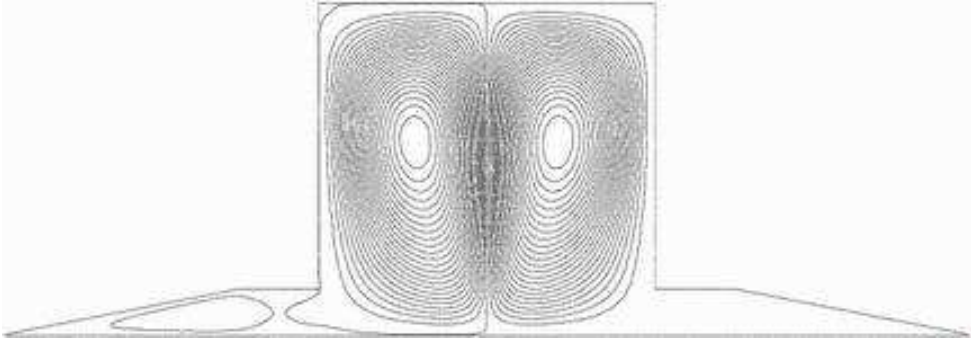}
\caption{\label{stabt} Two symmetrical
stationary Taylor vortices which are denoted by $\square$ on the
fig.\ref{tab1}. Here is  $\nu=0.8 (
10^{-4}\frac{m^2}{ s}),$ $w=0~,$ $u_{max}=0.07~,$
$\mbox{Re}=1570,~\mbox{Re}_{ax}=0.$}
\end{figure}

Let us consider the pure rotation case without any axial flow. The most probable state for a short cylinder with an aspect ratio $\Gamma
\sim 1$ can be described as follows. The nearly quadratic cross
section of the annulus is filled with two counter rotating stable
Taylor vortices as represented on the fig. \ref{stabt}.
We proved that this structure remains mirror symmetric for a large region
of azimuthal Reynolds numbers, $ \mbox{Re} \in [0,2500].$ The
investigated states are denoted in fig.\ref{tab1} by boxes, $\square$,
on the horizontal axes. All of them have quite the same structure
- two stationary Taylor vortices in the annulus up to very high
azimuthal Reynolds numbers.

\begin{figure}[htbp]
\center
\includegraphics[width=45mm, angle=270]{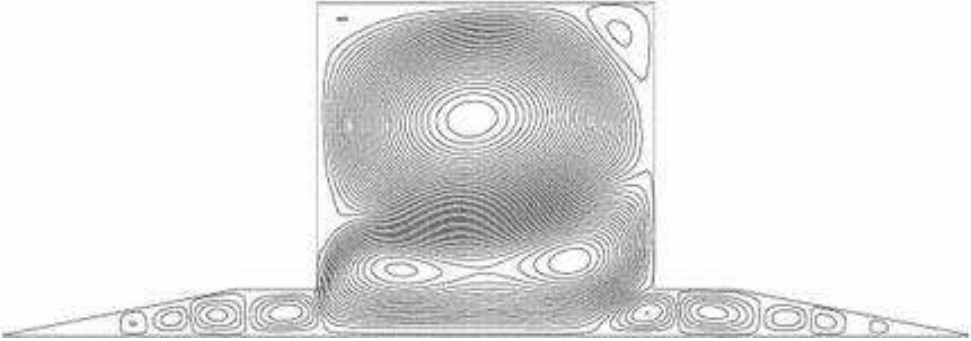}
\caption{\label{fine}  A lot of fine-grained structures of small vortices
  overlaid by large scale vortices. Here $\nu=0.1 (10^{-4}\frac{m^2}{ s}),$
 $w=0~,$ $ u_{max}=0.04~,$ $\mbox{Re}=12560~,$ $\mbox{Re}_{ax}=0.$}
\end{figure}

\begin{figure}[htbp]\unitlength1cm
\center
\includegraphics[width=45mm,angle=270]{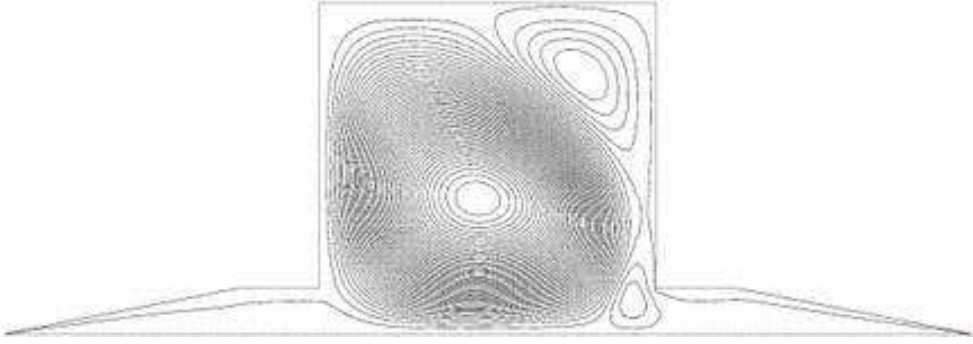}
\caption{\label{alone} One large
stationary vortex filling out nearly the whole annulus (denoted on the
fig.\ref{tab1} by {\Large$\bullet$}). Here $\nu=0,1 (10^{-3}\frac{m^2}{ s});$
$w=0,05~;$ $u_{max}=0,04~;$ $ \mbox{Re}=1256 ~;$ $ \mbox{Re}_{ax}=0,32.$}
\end{figure}

An enlargement of the azimuthal Reynolds numbers leads to the second
state in the annulus.  For very large azimuthal Reynolds numbers
$\mbox{Re} >4000$ the intrinsic instabilities give rise to perturbed
states with a lot of small vortices. This flow pattern can be
interpreted as a region with a fine-grained structure overlaid with a
large scale structure - three Taylor-like vortices (see
fig.~\ref{fine}).
Consider now the onset on an axial flow with relatively small
velocity.  The existence of the axial flow can be considered as a
small perturbation of the base flow with pure azimuthal velocity. In
the region of the azimuthal Reynolds numbers $\mbox{Re} \le 1500 $ the
axial flow promoted one of the counter rotating Taylor vortices in
appropriate direction. As result the annulus will be filled also for
very small axial amplitudes of the axial flow with one stable vortex
(see fig.~\ref{alone}). We see the abrupt change of the patterns
structure from two stable Taylor vortices to one by switching on an
axial flow. This is the third pattern structure observed in this
system. It is denoted by {\Large$ \bullet $} on the fig.\ref{tab1}.

\begin{figure}[ht]
\setlength{\unitlength}{1cm}
\begin{minipage}[t]{7cm}
\center
\includegraphics[width=2.4cm, angle=270]{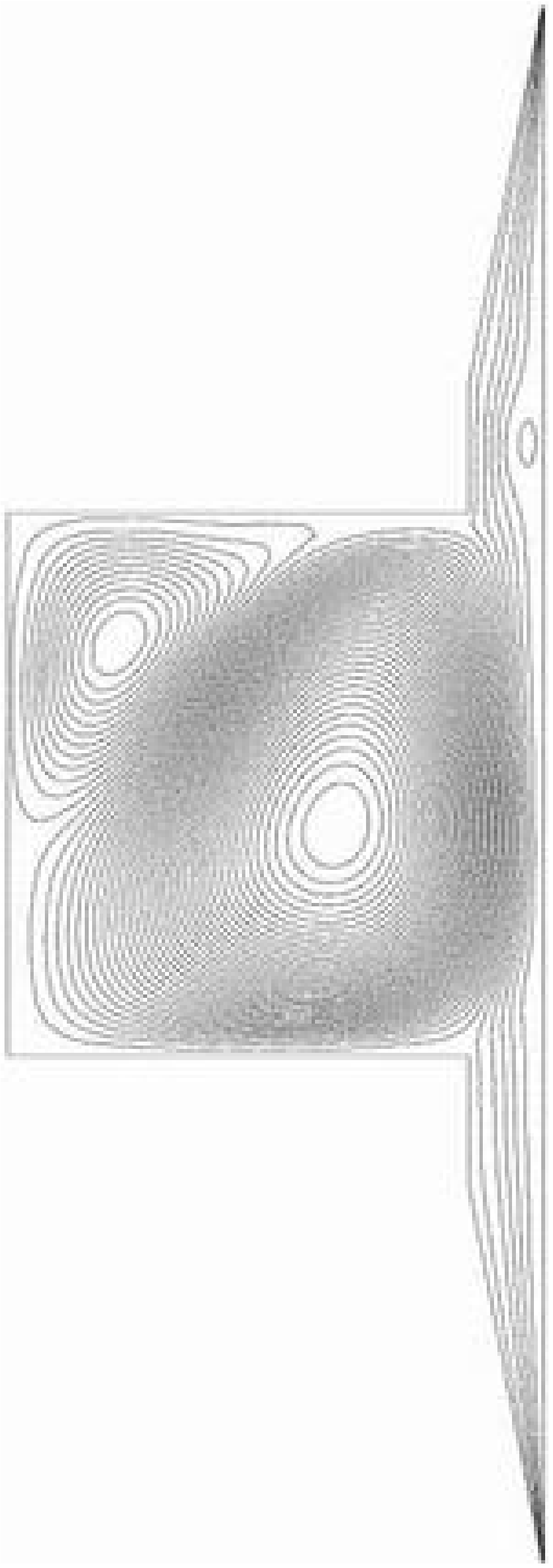}
\vspace{0.1cm}
\center a) $t_1=0.402$
\end{minipage}\hfill
\begin{minipage}[t]{7cm}
\center
\includegraphics[width=2.4cm, angle=270]{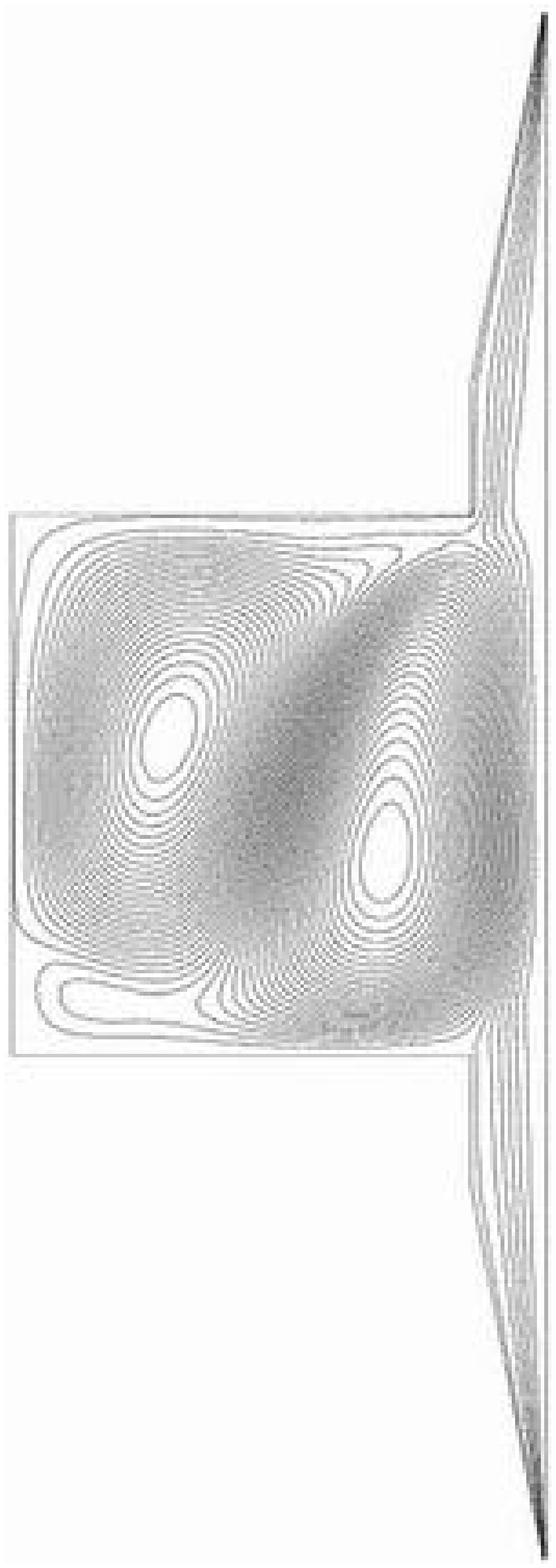}
\vspace{0.1cm}
\center b) $t_2=0.427$ (minimum energy)
\end{minipage}
\center
\vspace{0.3cm}
\begin{minipage}[t]{7cm}
\center
\includegraphics[width=2.4cm, angle=270]{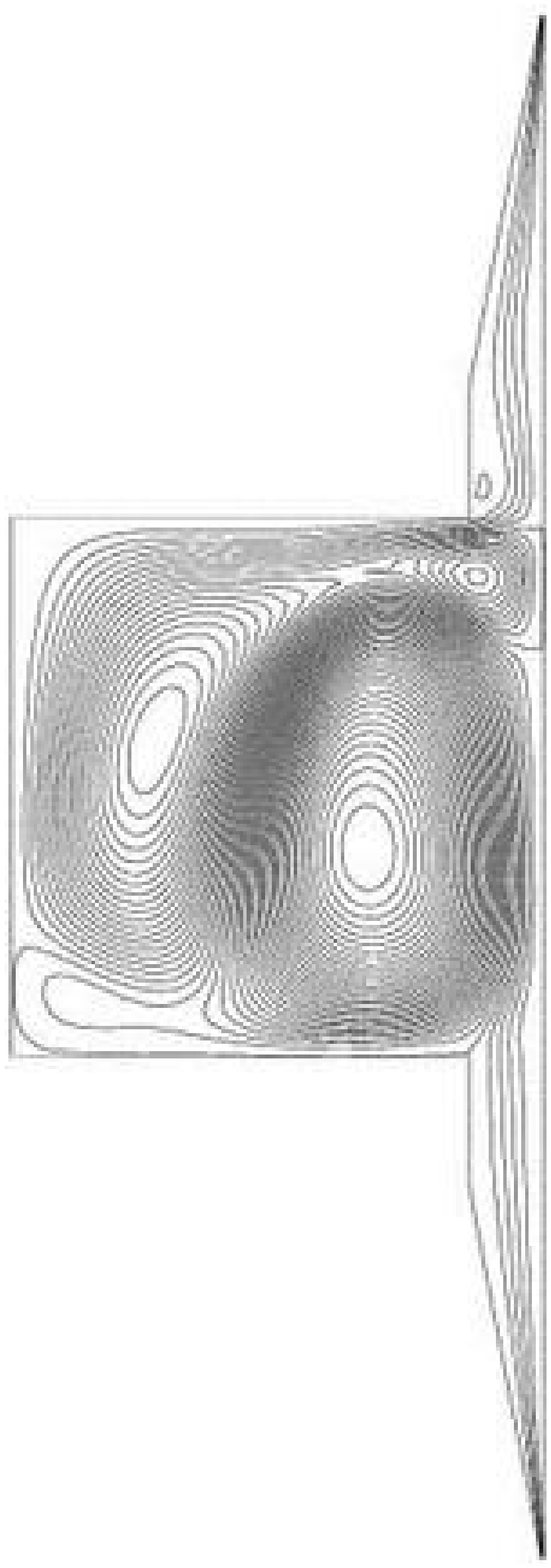}
\vspace{0.1cm}
\center c) $t_3=0.478$
\end{minipage}\hfill
\begin{minipage}[t]{7cm}
\center
\includegraphics[width=2.4cm, angle=270]{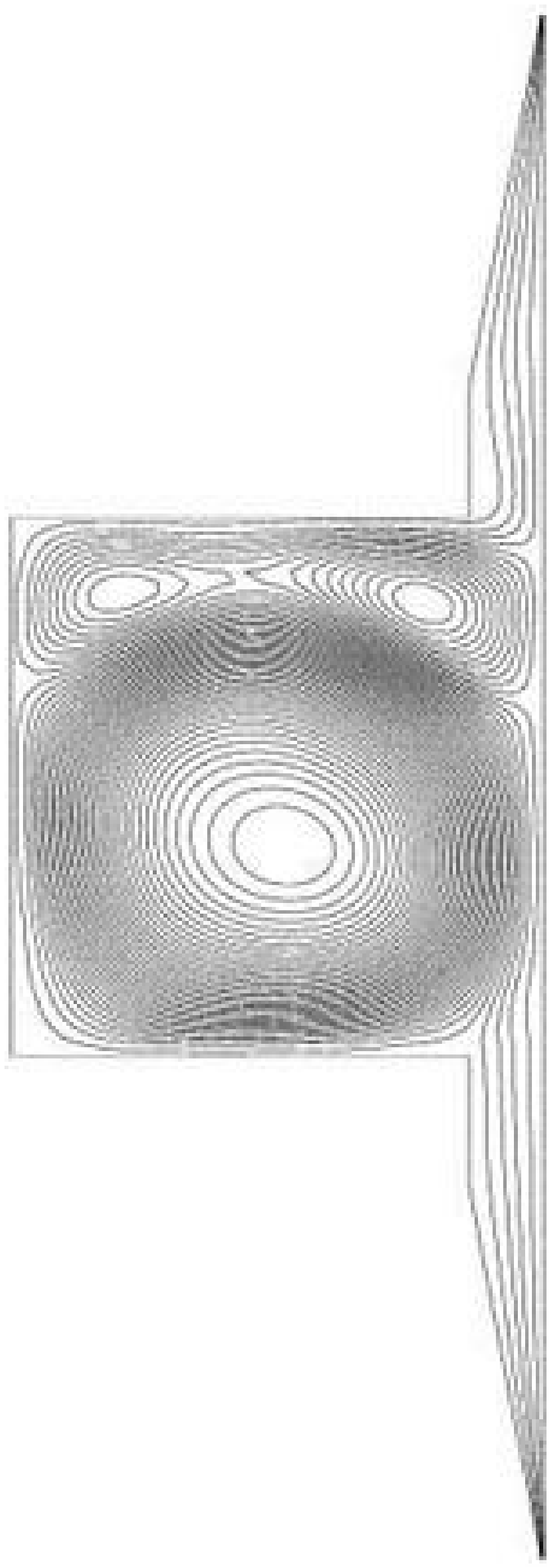}
\vspace{0.25cm}
\center d) $t_4=0.490$ (maximum energy)
\end{minipage}
\caption{\label{unsteady} The natural quasi periodic
oscillations of a two vortex state in the annulus (denoted on the 
fig.~\ref{tab1} by $\bigstar$) in  four different moments of
dimensionless time. Here is  $\nu=0.3 (10^{-4}\frac{m^2}{ s}),$ $w=0.30~,$ 
$ u_{max}=0.05~,$ $\mbox{Re}=5108 ~,$ $ \mbox{Re}_{ax}=6.333.$ }
\end{figure}

In the region on the Reynolds numbers $\mbox{Re}_{az} \sim 1500,$ $
\mbox{Re}_{ax} \sim 0,4 $ there exists a branch point in which three
possible flow states meet. For all axial Reynolds numbers $ 0.05
<\mbox{Re}_{ax}<4 $ and $ \mbox{Re}<1500 $ it is the first branch for
which is characterized by one stable vortex in the annulus
(fig.\ref{alone}).
For $1250<\mbox{Re}<2400$ and for the small axial Reynolds numbers
 $0<\mbox{Re}_{ax}<0.64$ a second branch with two stable Taylor
 vortices appears (fig.\ref{stabt}).  For $\mbox{Re}_{az}>1500$ and
 $0.3<\mbox{Re}_{ax}<0.9$ the third branch appears with two quasi
 periodically natural oscillating vortices (fig.\ref{unsteady}).
 
\begin{figure}[ht]
\center
\includegraphics[width=10cm]{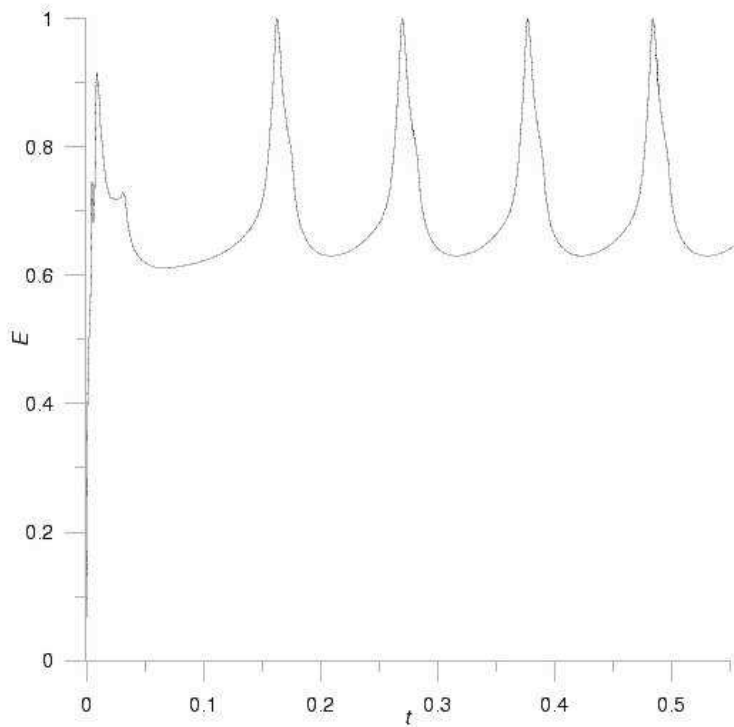}
\caption{\label{autooscil1} The time dependence of the total energy $E(t)$
for the flow by $\mbox{Re}=5108,$ $ \mbox{Re}_{ax}=6,333. $}
\end{figure}

\begin{figure}[ht]
\center
\includegraphics[width=10cm]{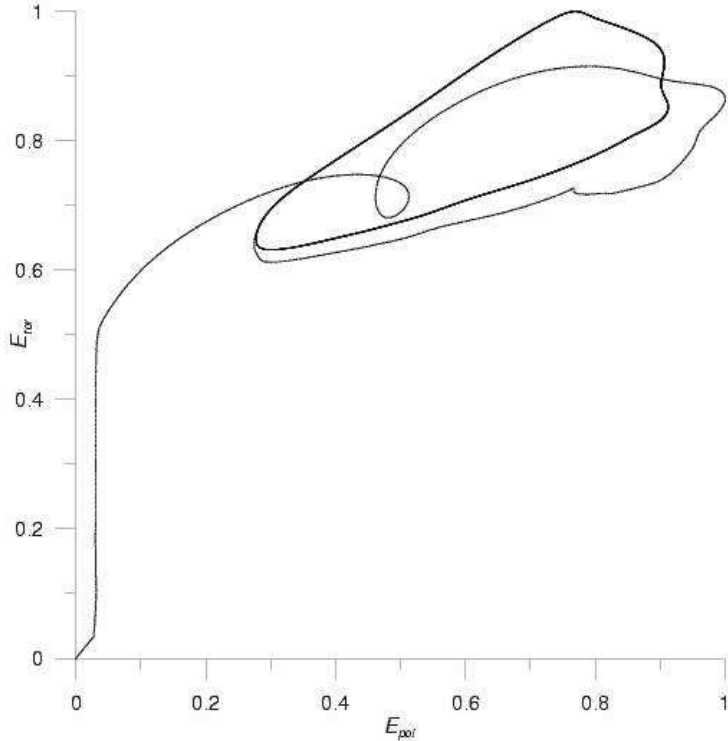}
\caption{\label{autooscil2} The parametric representation of the
  energy exchange between $E_{tor}$ and $E_{pol}$ in the plane of
 both energies by $\mbox{Re}=5108~,$ $\mbox{Re}_{ax}=6,333$.}
\end{figure}

This branch representes non stationary states. The energy and momentum
of the system show also typical quasi periodical oscillations.  The
time dependence of the energy is given on the fig.~\ref{autooscil1}.
The time dependence of the angular momentum is of the same type.  We
rendered the energy dimensionless using its value for small Reynolds
numbers, $\mbox{Re}=251.2, ~\mbox{Re}_{ax}=0.02 ~.$ In
fig.~\ref{autooscil1} we also see that maximal values for the
energy are smaller in the perturbation state as in the non perturbed
flow.
We will distinguish two different types of energy in the system - the
rotation or toroidal energy, defined by
$$E_{tor}=2\pi \int v ^2 r dr dz$$ and the energy caused by axial and
local radial flow in the system
$$E_{pol}=2\pi\int\left(u^2+w^2\right)r dr dz, $$ named also poloidal
energy.  Correspondingly with Reynolds numbers the toroidal energy is
three order larger as the last one. But the intrinsic instabilities in
the system lead to steady energy exchange between different motion
types. This exchange we demonstrate by a parametric representation in
the plane of both energies in fig.~\ref{autooscil2}.

The flow field between differently shaped bodies of revolution were
studied experimentally or theoretically rarely. The experimental work
devoted to the case which is somewhat analogously to the motion in the
conical slits at the caps of cylinders in our case was done by Wimmer
\cite{wimm}.  For a rotating cylinder in a stationary cone he proved
that due to the non constant gap different types of motion can coexist
in the gap. By the growth of the Reynolds numbers the first vortices
appear at the location of the largest gap size and are deformed,
having larger axial extension. The numerical studies in our geometry
confirm this statement as can be seen on the fig.~\ref{fine}.

\section{Conclusions}

\begin{figure}[htbp]
\center
\includegraphics[width=140mm]{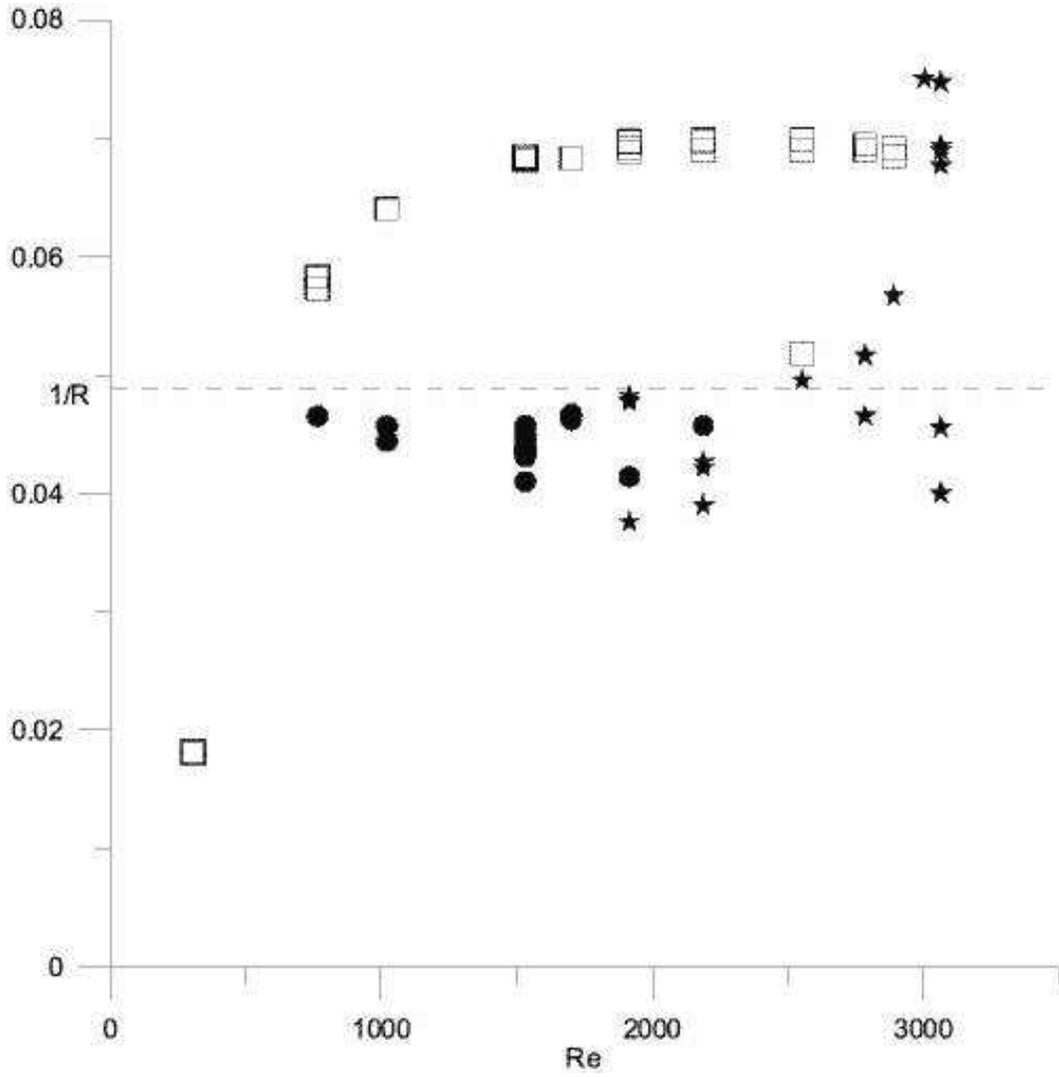}
\caption{\label{tab2} The distribution of the radial component of the 
dimensionless velocity $\max {\tilde u}$ in the Taylor-Coutte flow
for different states in dependence on
the Reynolds numbers $\mbox{Re}$ and $\mbox{Re}_{ax}.$ Two Taylors vortices
in the annulus are denoted by $\square$, one stable vortex by
{\Large$\bullet$}, natural oscillation state of two vortices by
$\bigstar .$}
\end{figure}

The convenient asymptotic representation for the exact steady solution
of the Taylor-Couette problem was found in the case of closed short
cylinders $ \Gamma \sim 1$ (section \ref{sero}).  The found
approximated solution (\ref{appsov}) will be an exact solution for
infinitely large viscosity or for $Re \to 0$.  For finite Reynolds
numbers a flow in a thin layer along the caps of the cylinder can
arise. We estimated the radial and axial components of its velocity.
At first blush in the case of the strong nonlinear motion the
estimation $u/V_0 \sim R^{-1/2} ,$ $w/V_0 \sim R^{-1/2} $ must hold.
This follows from the comparison of the nonlinear terms in the
Navier-Stokes equations. But the numerical computations for the short
cylinders with $R \gg 1$ and large Reynolds numbers lead to another
estimation, $u/V_0 \sim R^{-1} ,$ $w/V_0 \sim R^{-1}, $ which can be
obtained from the Navier-Stokes equations in the case of very large
viscosity.  In fig.~\ref{tab2} we represent the radial component of
the dimensionless velocity only. For the other component the pattern
distribution is quite the same.

From the estimations it follows that the azimuthal component of the
velocity has the main influence on the stability of the
system. Consequently, we can use for our qualitative stability
analysis the found approximated solution.

The stability analysis done in section \ref{loc} showed that the
inhomogeneity of the base flow in the radial direction as well as the
axial flow cause a perturbation of the Rossby waves type.  We proved
that the $\bf Z^2-$symmetry breaking due to the axial flow in the
Taylor-Couette flow is in the same time the helicity symmetry breaking
(section \ref{helica}).  The states with non-zero helicity can accrue
in the system (section \ref{resul}). The energy oscillations in the
case of the unsteady pattern with natural oscillations as in
fig.~\ref{autooscil1} correspond to the helicity oscillations in the
system.This can be seen from fig.~\ref{unsteady}.

The weak axial flow in an annulus between the rotating inner cylinder
and the fixed outer cylinder has several important engineering
applications in particular if the axial flow is directed as a thin
layer along the surface of the rotating cylinder.  We studied this
case and proved that the axial flow stabilizes the motion and that in
a system with very narrow annulus different super critical states are
possible (section \ref{resul}). We remark that experiments in short
cylinders with an aspect ratio of $\Gamma \sim 1$ without any axial
flow showed a poor variety of states \cite{furu}, \cite{wa}. But 
for smaller aspect
ratio all the states which we described in our case can be
observed. This behavior of the Taylor-Couette system can be
interpreted as follows. The influence of the weak axial flow along the
surface of the inner cylinder can be considered as a change in the
geometry, i.e., as a shortening of the axial length of the cylinder.
This interpretation is supported by simple estimations.

For technical applications it is important to describe not only the
velocity field but also the pressure distribution in the device. The
pressure differences in the annulus coming from the vortex motion can
be estimated as $R^{-1}$ of the pressure difference corresponding to
the axial flow.

\section*{Acknowledgments}
The authors are grateful M. V. Babich, S.K. Matveev, V. M. Ponomarev,
A.I. Shafarevich and G. Windisch for interesting and fruitful
discussions.  The work was kindly supported by the BMBF- project grant
number PIM3CB under the leadership of S. Pickenhain.

\end{document}